\documentclass[aps,groupedaddress,pre,twocolumn,showpacs,floatfix]{revtex4}
\usepackage[dvips]{graphicx}
\usepackage[dvips]{hyperref}
\usepackage{amsmath}
\usepackage{amsfonts}
\usepackage{amssymb}
\providecommand{\U}[1]{\protect\rule{.1in}{.1in}}
\hypersetup{colorlinks, citecolor=red, filecolor=blue,
urlcolor=blue, pdfpagemode=None, pdfsubject={Synchronization},
pdfauthor={Michael Cross}}
\begin{document}
\title{Synchronization by Reactive Coupling and Nonlinear Frequency Pulling}
\author{M.~C.~Cross}
\affiliation{Department of Physics 114-36, California Institute of Technology, Pasadena,
California 91125}
\author{J.~L.~Rogers}
\affiliation{HRL Laboratories, LLC, 3011 Malibu Canyon Road, Malibu, California 90265}
\author{Ron Lifshitz}
\affiliation{School of Physics \& Astronomy, Raymond and Beverly
Sackler Faculty of Exact Sciences, Tel Aviv University, Tel Aviv
69978, Israel.}
\author{A.~Zumdieck}
\affiliation{Max Planck Institute for Physics of Complex Systems, Noethnitzer Strasse 38,
01187 Dresden, Germany}
\date{\today}

\begin{abstract}
We present a detailed analysis of a model for the synchronization of nonlinear
oscillators due to reactive coupling and nonlinear frequency pulling. We study
the model for the mean field case of all-to-all coupling, deriving results for
the initial onset of synchronization as the coupling or nonlinearity increase,
and conditions for the existence of the completely synchronized state when all
the oscillators evolve with the same frequency. Explicit results are derived
for Lorentzian, triangular, and top-hat distributions of oscillator
frequencies. Numerical simulations are used to construct complete phase
diagrams for these distributions.

\end{abstract}

\maketitle

\section{Introduction}

Increasingly the collective behavior displayed by groups of
interacting dynamical components, each of which may be capable of
a full range of complex dynamics, is essential to understanding
and ultimately designing systems.  Examples in biology, physics,
and engineering are diverse, ranging from understanding sensory
perception to the design of antennas capable of simultaneously
sending and receiving signals at the same frequency. While in
general the dynamical components may be functionally distinct and
heterogeneous, in many examples one is interested in the
collective behavior displayed by a group of similar coupled
elements.  One commonly studied example is the synchronization of
oscillating subsystems that interact.

The ability of a group of coupled oscillators to synchronize
despite a distribution in individual frequencies is a broadly
applicable phenomenon.  In physical systems coherent oscillations
can be used to enhance detector sensitivity or increase the
intensity of a power source.  Synchronization is also important in
biological systems where the collective behaviors in populations
of animals, such as the synchronized flashing of fireflies or the
coherent oscillations observed in the brain, serve as examples.
Including a distribution in individual frequencies in an otherwise
homogeneous ensemble captures some of the inevitable differences
that group members will possess.

Although synchronization is often put forward as an example of the
importance of understanding nonlinear phenomena, the intuition for
it, and indeed the subsequent mathematical discussion, often
reduces to simple linear ideas. For example, the famous example of
Huygens' clocks \cite{BSRW02} can be understood in terms of a
linear coupling of the two pendulums through the common mounting
support. Then the larger damping of the symmetric mode (coming
from the larger, dissipative motion of the common support)
compared with the antisymmetric mode leads at long times to a
synchronized state of the two pendulums oscillating in antiphase.
The nonlinearity in the system is simply present in the individual
motion of each pendulum; specifically in the mechanism to sustain
the oscillations. Without the drive, the oscillators would still
become synchronized through the faster decay of the even mode,
albeit in a slowly decaying state. Rather than this mode-dependent
dissipation mechanism, one might expect synchronization to arise
from the intrinsically nonlinear effect of the frequency pulling
of one oscillator by another. Furthermore, the model describing
the two pendulums, as well as most other models used to show
synchronization, has dissipative coupling between the oscillators.
In contrast, many physical situations have mainly reactive
coupling.

Nanoscale mechanical oscillator arrays are one example that offer significant
potential in a range of technologies \cite{BR02},\cite{SHSICC03}. Due to the
scales and numbers in these arrays, active control of individual oscillators
poses a number of challenging problems. Synchronization offers a potentially
appealing alternative in some applications. One notable characteristic of
these arrays is the predominantly reactive coupling due to elastic or
electrostatic interactions. This context is the motivation for our work. In
particular, we study the model%
\begin{equation}
\dot{z}_{m}=i(\omega_{m}-\alpha|z_{m}|^{2})z_{m}+(1-\left\vert z_{m}%
\right\vert ^{2})z_{m}+\frac{K+i\beta}{N}\sum_{n=1}^{N}(z_{n}-z_{m}),
\label{eq_cross}%
\end{equation}
where $z_{m}$ is a complex number representing the amplitude
$r_{m}$ and phase $\theta_{m}$ of the mth oscillator
$z_{m}=r_{m}e^{i\theta_{m}}$. The natural frequency $\omega_{m}$
of each oscillator is chosen from a distribution $g(\omega)$. We
call the width of the distribution $w$. Coefficients of the terms
for nonlinear frequency pulling ($\alpha$), dissipative coupling
($K$), and reactive coupling ($\beta$) serve as model parameters.
The coupling is taken to be infinite range or all-to-all.

In this paper we study the roles of nonlinear frequency pulling
and reactive coupling in the absence of the dissipative coupling
using Eq.~(\ref{eq_cross}) with $\alpha\neq0,\beta\neq0,K=0$. In
addition the distribution $g(\omega)$ of the $\omega_{m}$ must be
specified. We study the case of positive $\alpha$ and $\beta$; for
a symmetric distribution $g(\omega)$ the results are the same
changing the sign of both $\alpha$ and $\beta$.

Alternatively, when only nonlinear saturation and dissipative coupling are
present ($\alpha=\beta=0,K\neq0$) Eq.~(\ref{eq_cross}) reduces to
\begin{equation}
\dot{z}_{m}=(i\omega_{m}+1-\left\vert z_{m}\right\vert ^{2})z_{m}+\frac{K}%
{N}\sum_{n=1}^{N}(z_{n}-z_{m}).\label{AmplitudePhase}%
\end{equation}
The behavior for general $w$ and $K$ of Eq.~(\ref{AmplitudePhase})
has been analyzed by Matthews et al.\ \cite{MMS91}. If the width
$w$ of the distribution $g(\omega)$ is narrow, so that the time
evolution of the magnitudes $\left\vert z_{m}\right\vert $ is fast
compared with that of the phase dispersion, and $K$ is small,
$\left\vert z_{m}\right\vert $ rapidly relaxes to a value close to
unity, and the only remaining variable for each array member is
the phase $\theta_{m}$. Equation~(\ref{AmplitudePhase}) can then
be reduced to a simple form \cite{W67},\cite{K75}, often known as
the Kuramoto model
\begin{equation}
\dot{\theta}_{m}=\omega_{m}+\frac{K}{N}\sum_{n=1}^{N}\sin(\theta_{n}%
-\theta_{m}),\label{Kuramoto}%
\end{equation}
that has been the subject of numerous studies \cite{ABVRS05}. In the absence
of coupling each oscillator in this model would simply advance at a rate that
is constant in time, but with some dispersion of frequencies over the
different elements. Equation~(\ref{Kuramoto}) is an abstraction from the
equations describing most real oscillator systems, leaving out many important
physical features.

The complex notation of Eq.~(\ref{eq_cross}) suggests the introduction of a
complex order parameter $\psi$ to measure the coherence of the oscillations%
\begin{equation}
\psi=R\,e^{i\Theta}=\frac{1}{N}\sum_{n=1}^{N}r_{n}e^{i\theta_{n}},
\label{OrderParameter}%
\end{equation}
with $r_{n}=1$ for the Kuramoto model, and then the equation of motion can be
written [take the imaginary part of $\psi e^{-i\theta_{m}}$ in
Eq.~(\ref{OrderParameter}) to evaluate the sum appearing in
Eq.~(\ref{Kuramoto})] as%
\begin{equation}
\dot{\theta}_{m}=\omega_{m}+KR\sin(\Theta-\theta_{m}). \label{KuramotoField}%
\end{equation}
Thus the behavior of each oscillator is given by its tendency to lock to the
phase of the order parameter. The term $KR\sin(\Theta-\theta_{m})$ acts as a
locking force, and locking occurs for all oscillators with frequencies
satisfying $\left\vert \omega_{m}\right\vert <KR$, with the locked oscillator
phase given by $\Theta+\sin^{-1}(\omega_{m}/KR)$. The magnitude $R$ of the
order parameter must then be determined self-consistently via
Eq.~(\ref{OrderParameter}). The generalization of the locking force to apply
to the model Eq.~(\ref{eq_cross}) will be a conspicuous feature of our work.

Equation~(\ref{Kuramoto}) is known to show rich behavior,
including, in the large $N$ limit, a sharp synchronization
transition at some value of the coupling constant $K=K_{c}$
\cite{K75}, which depends on the frequency distribution of the
uncoupled oscillators $g(\omega)$. The transition is from an
unsynchronized state with $\psi=0$ in which the oscillators run at
their individual frequencies, to a synchronized state with
$\psi\neq0$ in which a finite fraction of the oscillators lock to
a single frequency. The transition at $K_{c}$ has many of the
features of a second order phase transition, with universal power
laws and critical slowing down \cite{K75}, as well as a diverging
response to an applied force \cite{S88}.

We present a detailed analysis of the model Eq.~(\ref{eq_cross})
with $K=0$ describing the synchronization of an oscillator
ensemble involving reactive coupling between the elements, which
then leads to synchronization through nonlinear frequency pulling.
We begin in Sec.~\ref{sec_derive} by deriving Eq.~(\ref{eq_cross})
as a description of arrays of nanoelectromechanical oscillators.
We then discuss the solutions to Eq.~(\ref{eq_cross}) for a
variety of symmetric distributions in intrinsic frequencies
$g(\omega)$. Common solution types exist for the three distinct
types of frequency distributions we studied. In
Sec.~\ref{sec_examples} we introduce these solutions and the
measures we use to describe them. We begin the analysis in
Sec.~\ref{sec_form} by moving to a continuum description to derive
synchronization conditions. We are able to analyze the existence
of two behavior types in closed form: the unsynchronized solution
in Sec.~\ref{sec_onset} and the fully locked synchronized behavior
in Sec.~\ref{Sec_FullLocking}. Results from our analytical
arguments are combined with those from simulations to present the
solutions and associated phase diagram for each frequency
distribution in Sec.~\ref{sec_results}.  We begin these by
considering an unbounded symmetric frequency distribution in the
form of a Lorentzian in Sec.~\ref{sec_lort}.  Continuing in this
section we then approximately double the distribution width to
discuss some interesting characteristics of the well-ordered
synchronized solutions as well as changes to the bifurcations. In
contrast, Sec.~\ref{sec_top} presents the results for a bounded
distribution when all frequencies over some range are equally
likely (a top-hat distribution). This is followed in
Sec.~\ref{sec_tri} by the case of a symmetric unimodal frequency
spread, namely a triangular distribution. Finally, conclusions are
made in Sec.~\ref{sec_conclude}.

A brief account of some results for the model Eq.~(\ref{eq_cross})
has been reported previously \cite{CZLR04}.

\section{Connection with Mechanical Oscillators}

\label{sec_derive} Although the main focus of this paper is analyzing the
behavior of Eq.~(\ref{eq_cross}) we first show how such an equation might
arise from the equations of motion of physical oscillators. As an example
consider the system defined by%
\begin{equation}
\ddot{x}_{n}+\omega_{n}^{2}x_{n}-\nu(1-x_{n}^{2})\dot{x}_{n}+ax_{n}%
^{3}-D[x_{n}-\tfrac{1}{2}(x_{n+1}+x_{n-1})]=0. \label{Eq_DV}%
\end{equation}
The first two terms describe uncoupled harmonic oscillators. We suppose the
uncoupled oscillators have a linear frequency that is near unity by an
appropriate scaling of the time variable%
\begin{equation}
\omega_{n}^{2}=1+\delta_{n},\quad\delta_{n}\ll1.
\end{equation} The third term is a \emph{negative} linear damping,
which represents an energy source to sustain the oscillations, and
positive nonlinear damping, so that the oscillation amplitude
saturates at a finite value. Again this saturation value is chosen
to be of order unity by an appropriate scaling of the
displacements $x_{n}$. For an example of an effective negative
linear damping term in a micromechanical oscillator see references
\cite{ZZOTSICPC01, IKARC05}. One could also imagine implementing
such an effect with an electronic feedback loop sensing each
oscillator velocity and driving the oscillator with an appropriate
phase. The first three terms of Eq.~(\ref{Eq_DV}) comprise a set
of uncoupled van der Pohl oscillators. The term $ax_{n}^{3}$
describes a stiffening of the spring constant (for $a>0$) and is a
reactive nonlinear term that leads to an amplitude dependent shift
of the resonant frequency. With $\nu=0$ this would give us
\emph{Duffing} oscillators. The final term is a nearest neighbor
coupling between the oscillators, depending on the difference of
the displacements. This is a reactive term, that conserves the
energy of the system. Others \cite{AEK90} have considered
nonlinear oscillators coupled through their velocities; this is a
dissipative coupling that would lead to $K\neq0$ in the
amplitude-phase reduction.

The complex amplitude equation (\ref{eq_cross}) holds if the
parameters $\nu,\alpha,D,\delta_{n}$ are sufficiently small. In
this case the equations of motion are dominated by the terms
describing simple harmonic oscillators at frequency one, and the
time dependence remains close to $e^{\pm it}$. To formalize the
smallness of $\nu,\alpha,D,\delta_{n}$ we introduce a small
parameter $\varepsilon$ and write $\nu=\varepsilon\bar{\nu}$,
$\delta _{n}=\varepsilon\bar{\delta}_{n}$, $a=\varepsilon\bar{a}$,
$D=\varepsilon \bar{D}$. The oscillating displacement is written
as a slow modulation of oscillations at frequency one, plus
corrections \cite{LC03}%
\begin{equation}
x_{n}(t)=\left[  A_{n}(T)e^{it}+c.c.\right]  +\epsilon x_{n}^{(1)}%
(t)+\ldots\label{ansatz1}%
\end{equation}
with $T=\varepsilon t$ a slow time scale \footnote{The reader
might be familiar with similar calculations where the amplitude of
oscillations is small so that
$x_{n}\rightarrow\varepsilon^{1/2}A_{n}e^{it}+c.c.$, as in
ref.~\cite{LC03}. However in the present case inspection of the
van der Pohl term shows that the oscillations will grow until
$x_{n}=O(1)$. The nonlinear effects are small because the
multiplying coefficients $\nu$ and $a$ are small. Of course, this
difference results from a choice of scaling of the $x_{n}$
variables; for example choosing $\bar{x}_{n}=\nu^{1/2}x_{n}$ would
result in the more familiar procedure.}. The variation of
$A_{n}(T)$ gives us the extra freedom to eliminate secular terms
and ensure that the perturbative correction $x_{n}^{(1)}(t)$, as
well as all higher-order corrections to the linear response, do
not diverge (as they do if one uses naive perturbation theory).
Using the relation
\begin{equation}
\dot{A}_{n}={\frac{dA_{n}}{dt}}=\epsilon{\frac{dA_{n}}{dT}}\equiv\epsilon
A_{n}^{\prime}, \label{eq:adot}%
\end{equation}
(denoting a time derivative with respect to the slow time $T$ by a prime) we
calculate the time derivatives of the trial solution~(\ref{ansatz1})
\begin{subequations}
\label{eq:derivs}%
\begin{align}
\dot{x}_{n}  &  =\left(  [iA_{n}+\epsilon A_{n}^{\prime}]e^{it}+c.c.\right)
+\epsilon\dot{x}_{n}^{(1)}(t)+\ldots\\
\ddot{x}_{n}  &  =\left(  [-A_{n}+2i\epsilon A_{n}^{\prime}+\epsilon^{2}%
A_{n}^{\prime\prime}]e^{it}+c.c.\right)  +\epsilon\ddot{x}_{n}^{(1)}(t)+\ldots
\end{align}
Substituting these expressions back into the equation of motion~(\ref{Eq_DV}),
and picking out all terms of order $\epsilon$, we get the following equation
for the first perturbative correction%
\end{subequations}
\begin{widetext}
\begin{align}
\ddot{x}_{n}^{(1)}+x_{n}^{(1)}=-\bar{\delta}_{n}A_{n}  &  -\left(
2iA_{n}^{\prime}e^{it}+c.c.\right)  +\bar{\nu}\left(  iA_{n}e^{it}%
+c.c.\right)  (1-\left(  A_{n}e^{it}+c.c.\right)  ^{2})\nonumber\\
&  -\bar{a}\left(  A_{n}e^{it}+c.c.\right)  ^{3}-\bar{D}[\left(  A_{n}%
-\tfrac{1}{2}(A_{n}+A_{n-1})\right)  e^{it}+c.c.]. \label{eq_xone}%
\end{align}
\end{widetext}
The collection of terms proportional to $e^{it}$ on the right-hand
side of Eq.~(\ref{eq_xone}), called the secular terms, act like a
force driving the simple harmonic oscillator on the left-hand side
at its resonance frequency. The sum of all the secular terms must
vanish so that the perturbative correction $x_{n}^{(1)}(t)$ will
not diverge. (Terms varying as $e^{\pm3it}$ contribute a finite
response to $x_{n}^{(1)}$.) This gives us an equation for
determining the slowly varying amplitudes $A_{n}(T)$%
\begin{widetext}
\begin{equation}
2A_{n}^{\prime}=(\bar{\nu}+i\bar{\delta}_{n})A_{n}-(\bar{\nu}-3i\bar
{a})\left\vert A_{n}\right\vert ^{2}A_{n}-i\bar{D}[A_{n}-\tfrac{1}{2}%
(A_{n+1}+A_{n-1})]=0. \label{DuffingAmpA}%
\end{equation}

More informally, we might write $x_{n}(t)=z_{n}(t)e^{it}+c.c+\cdots$ so that
$A_{n}(T)\rightarrow z_{n}(t)$ and Eq.~(\ref{DuffingAmpA}) can be written in
terms of the original \textquotedblleft small\textquotedblright\ parameters
without the formal scaling%
\begin{equation}
2\dot{z}_{n}=(\nu+i\delta_{n})z_{n}-(\nu-3ia)\left\vert z_{n}\right\vert
^{2}z_{n}-iD[z_{n}-\tfrac{1}{2}(z_{n+1}+z_{n-1})]=0 \label{DuffingAmp}%
\end{equation}
\end{widetext}
With a rescaling of time $\bar{t}=\nu t/2$ Eq.~(\ref{DuffingAmp})
reduces to the form Eq.~(\ref{eq_cross}) except that in
Eq.~(\ref{eq_cross}) the nearest neighbor coupling is replaced by
the all-to-all coupling convenient for theoretical analysis.

\section{Examples of Dynamical States}

\label{sec_examples} In this section we introduce the types of
dynamical states encountered for the model Eq.~(\ref{eq_cross}) as
well as diagnostic tools to characterize these states. The types
of states we find for the different distributions investigated are
largely the same, and so we use a particular example---a
Lorentzian distribution of oscillator frequencies with some
convenient choice of parameters $\alpha$ and $\beta$---and discuss
how the behavior depends systematically on the distribution and
other parameters later in the paper. We are mainly interested in
the behavior for large numbers of oscillators
$N\rightarrow\infty$. For the numerical simulations we are of
course restricted to finite $N$ (we typically use $N=1000$, but
have gone up to $N=100,000$ to investigate some subtle effects).
In our discussion we focus on those results that we expect to be
largely independent of $N$ for large $N$.

A key diagnostic for synchronization is the complex order
parameter $\Psi(t)$ defined by Eq.~(\ref{OrderParameter}),
introduced by Kuramoto \cite{K75}, with magnitude $R$ and phase
$\Theta$ where $r_{n}e^{i\theta_{n}}=z_{n}$. In the large $N$
limit, we could use a nonzero value of the order parameter at each
time as the basic criterion for a synchronized state. A
synchronized periodic state with frequency $\Omega$ would then
have $\Psi(t)=R\,e^{i\Omega t}$ with $R,\Omega$ constants. More
complicated dynamical states might also occur. For a finite number
of oscillators $N$ a precise diagnostic is harder, since there are
fluctuations of order $N^{-1/2}$ that make the instantaneous
$\Psi$ nonzero even in an unsynchronized state. For a synchronized
periodic state we could require that the time average
$\left\langle e^{-i\Omega(N)t}\Psi\right\rangle _{t}$
scales as $N$ as the number of oscillators changes for some
frequency $\Omega(N)$ that becomes constant for large enough $N$,
and use this as a measure of the synchronization. In practice we
use the simpler criterion that the time averaged magnitude
$\left\langle R(t)\right\rangle
_{t}=\left\langle|\psi(t)|\right\rangle _t$ is nonzero and does
not appear to decrease to zero as $N$ increases. (A $\Psi$
fluctuating about zero with an amplitude of order $N^{-1/2}$ will
of course lead to a nonzero $\left\langle R(t)\right\rangle _{t}$
of this same order.) This definition can also be applied to
aperiodic synchronized states.

Another useful diagnostic uses the actual frequency of each oscillator defined
by%
\begin{equation}
\tilde{\omega}_{n}=\frac{1}{t}\left[  \theta_{n}(t_{0}+t)-\theta_{n}%
(t_{0})\right],  \label{FrequencyLocking}%
\end{equation}
where $t$ is some long averaging time and $t_{0}$ is a starting
time sufficient to eliminate transients. A \emph{frequency locked}
state has a fraction of oscillators with the same frequency
$\tilde{\omega}_{n}$. (The fraction should be $O(1)$ and a value
not decreasing to zero as $N$ increases). If not all the
oscillators have the same frequency (i.e.\ the fraction is not
unity) we call the state \emph{partially locked}. If all the
oscillators have the same average frequency, we call the state
\emph{fully locked}. To make contact with the analytic results we
actually use a stricter criterion, and also require the magnitudes
$r_{n}$ to be time independent in the fully locked state. Usually
we find that a nonzero $\left\langle R(t)\right\rangle _{t}$ is
associated with frequency locking, but this is not always the
case.

\emph{Phase locking} is a stricter requirement than
Eq.~(\ref{FrequencyLocking}). For phase locking we would require
that $\theta_{n}(t)-\theta_{m}(t)$ does not diverge as
$t\rightarrow\infty$ for $m,n$ taken from some finite fraction of
the oscillators. (Frequency locking is, for example, consistent
with phase differences that grow diffusively proportional to
$t^{1/2}$.) We do not investigate this stricter locking criterion.

To investigate whether each locked oscillator in a frequency locked state is
tightly locked in phase to the phase of the order parameter, or is fluctuating
about this value, we define a \emph{phase synchronization index} for each
oscillator%
\begin{equation}
\chi_{n}=1-\frac{1}{2}\left\langle \left\vert e^{i\bar{\theta}_{n}(t)}%
-\frac{\left\langle e^{i\bar{\theta}_{n}(t)}\right\rangle _{t}}{\left\vert
\left\langle e^{i\bar{\theta}_{n}(t)}\right\rangle _{t}\right\vert
}\right\vert ^{2}\right\rangle _{t}, \label{eq_sync_ind}%
\end{equation}
with $\bar{\theta}_{n}$ the phase of the $n$th oscillator relative to the
order parameter phase%
\begin{equation}
\bar{\theta}_{n}(t)=\theta_{n}(t)-\Theta(t).
\end{equation}
When the phase of oscillator $n$ is locked to the order parameter
one, $\bar{\theta}_{n}(t)$ is constant and the phase
synchronization index is unity, whereas as $\bar{\theta}_{n}(t)$
tends towards a uniform distribution from $0$ to $2\pi$ the index
approaches zero. We define the number of oscillators with
$\chi_{n}$ very close to one as the \emph{tightly locked
cluster}.%

\begin{figure}
[tbh]
\begin{center}
\includegraphics[height=2.9473in,width=3.3in]{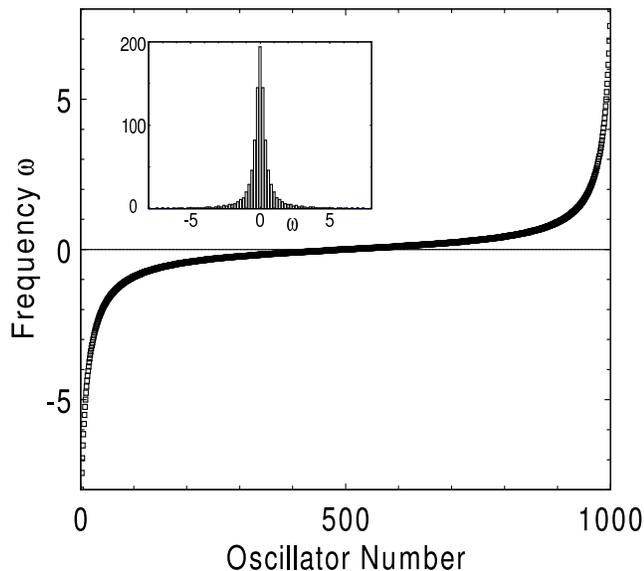}
\caption{Distribution $g(\omega)$ of frequencies for cutoff Lorentzian with
$g(0)=1$ and cutoff $\omega_{c}=8$ for $N=1000$ oscillators. The main graph
shows the individual frequencies, and the inset a histogram.}%
\label{Fig_lor_dist}%
\end{center}
\end{figure}
We now show results of numerical simulations for the Lorentzian distribution
of frequencies. We have chosen to cut off the distribution at some large
$\omega_{c}$ otherwise there would be a few very fast oscillators restricting
the time stepping of the numerics. Thus we use%
\begin{equation}
g(\omega)=\left\{
\begin{tabular}
[c]{ll}%
$g(0)\frac{w^{2}}{\omega^{2}+w^{2}}$ & for $\left\vert \omega\right\vert
<\omega_{c}$\\
$0$ & for $\left\vert \omega\right\vert >\omega_{c}$%
\end{tabular}
\ \ \ \ \ \ \ \ \ \ \right.  \label{eq_para_lort}%
\end{equation}
for a given choice of $g(0)$, with $w$ then fixed by the normalization
condition%
\begin{equation}
2wg(0)\tan^{-1}(\omega_{c}/w)=1\text{.}%
\end{equation}
(It is useful to parameterize the distribution in terms of $g(0)$,
since this quantity determines the values of $\alpha$ and $\beta$
at which synchronization occurs in the large $\alpha\beta$ limit
where the model reduces to the Kuramoto phase model
Eq.~(\ref{Kuramoto}).) In presenting the results we choose a
distribution width such that $g(0)=1$. For a cutoff $\omega_{c}=8$
this gives $w\simeq(f\pi)^{-1}$, with $f=0.974$. With no cutoff,
we would have $f=1$. The distribution of frequencies is generated
from
a uniform distribution of $N$ values $v_{n}$ on the interval $-\frac{1}%
{2}<v_{n}<\frac{1}{2}$ using the transformation%
\begin{equation}
\omega_{n}=w\tan(\pi fv_{n}).
\end{equation}
We present results for a deterministic set of $\nu_{n}$%
\begin{equation}
\nu_{n}=\left[  n-\frac{1}{2}(N+1)\right]  /(N-1),
\end{equation}
but have also used $\nu_{n}$ generated randomly on the interval. The
distribution of frequencies for $N=1000$ oscillators and a cutoff $\omega
_{c}=8$ is shown in Fig.\ \ref{Fig_lor_dist}.%

\begin{figure}
[tbh]
\begin{center}
\includegraphics[width=3.3in]{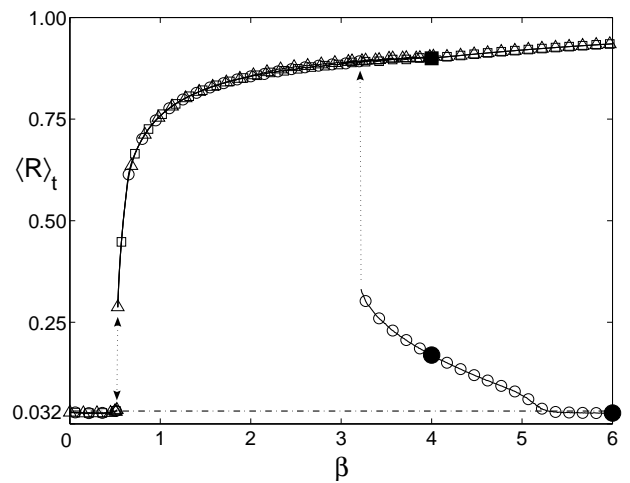}%
\caption{Solutions observed in simulations of $N=1000$ oscillators
with the cutoff Lorentzian distribution of intrinsic frequencies
of Fig.~\ref{Fig_lor_dist}. Shown is the time averaged order
parameter magnitude $\langle R\rangle_{t}$ over a range of $\beta$
values at constant $\alpha=1.5$. Solid lines are the observed
solutions. The overlapping symbols are representative results from
simulations following different solution branches with increasing
$\beta$ ($\triangle$) and decreasing $\beta$ ($\Box$ and
$\bigcirc$). Time traces associated with the solid symbols are
shown in Fig.~\ref{Fig_Orderparameters}. The dot-dashed line is at
a value of $\langle R\rangle_{t}=N^{-1/2}=0.032$, the order of
magnitude expected for random
fluctuations of the order parameter about zero.}%
\label{Fig_Lorentzian_Cut}%
\end{center}
\end{figure}
A plot of the dependence of the mean magnitude of the order parameter
$\left\langle R\right\rangle _{t}$ as a function of $\beta$ for fixed value of
$\alpha=1.5$ and $N=1000$ is shown in Fig.~\ref{Fig_Lorentzian_Cut}. This plot
shows three types of states: an unsynchronized state with $\left\langle
R\right\rangle _{t}$ essentially zero; and two synchronized states, one with
small $\left\langle R\right\rangle _{t}$, growing from $0$ to about $0.3$ as
$\beta$ decreases below about $5.2$, and one with large $\left\langle
R\right\rangle _{t}$ that exists for all $\beta>0.6$. (We will discuss later
such issues as whether the growth of the large $\left\langle R\right\rangle
_{t}$ is continuous or discontinuous near $\beta=0.6$.)%

\begin{figure}
[tbh]
\begin{center}
\includegraphics[width=3.3in]{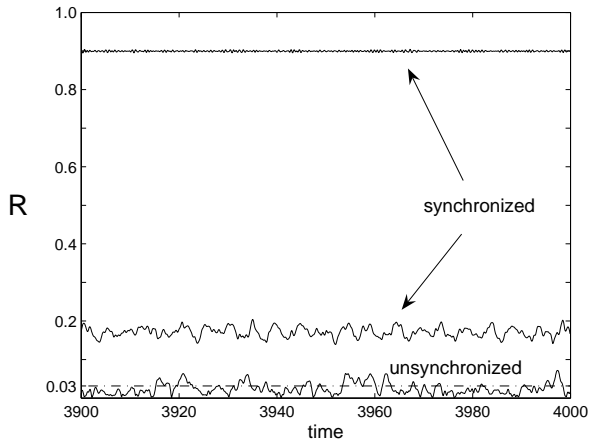}%
\caption{Order  parameter magnitude $R$ as a function of time
observed at the solid symbols in Fig. 2. The three traces are for
the unsynchronized state at $\beta=6.0$ (lower trace), and the
small amplitude synchronized state (middle trace) and
large amplitude synchronized state (upper trace) both at $\beta=4.0$.}%
\label{Fig_Orderparameters}%
\end{center}
\end{figure}
The variation of $R(t)$ for representative examples of these three states is
shown in Fig.~\ref{Fig_Orderparameters}. The lower trace shows the
unsynchronized state at $\beta=6.0$ in Fig.~\ref{Fig_Lorentzian_Cut}. The
fluctuations in $R(t)$ are consistent with fluctuations of the order parameter
about zero with magnitude of order $N^{-1/2}$ as expected for a collection of
$N$ oscillators with different frequencies and random phases. The average
frequency distribution $\tilde{\omega}_{n}$ (not shown) is unchanged from the
bare distribution $\omega_{n}$ shifted by $\alpha+\beta$. This shift can be
understood as arising from the nonlinear frequency shift with $\left\vert
z_{n}\right\vert =1$, and the coupling to a distribution of oscillators with
$\left\vert z_{m}\right\vert =1$ and random phases.%

\begin{figure}[tbh]
\begin{center}
\includegraphics[width=3.0in]{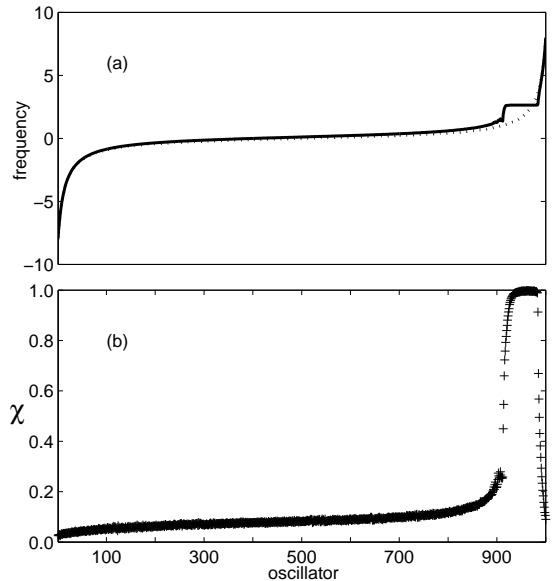}%
\caption{Frequency distribution and synchronization index for the small
amplitude synchronized state in Fig.~\ref{Fig_Orderparameters}: (a) -
actual frequency distribution (solid line) and bare frequency distribution
shifted by $\alpha+\beta$ (dotted line); (b) - synchronization index
$\chi_{n}$.}%
\label{Fig_smr_plateau}%
\end{center}
\end{figure}
The state corresponding to the middle trace in Fig.~\ref{Fig_Orderparameters}
is the low amplitude state at $\beta=4.0$ in Fig.~\ref{Fig_Lorentzian_Cut}.
The mean order parameter magnitude $\left\langle R\right\rangle _{t}%
\simeq0.17$ is much larger than $N^{-1/2}\simeq.032$, suggesting this is a
synchronized state. The corresponding frequency distribution and
synchronization index are shown in Fig.~\ref{Fig_smr_plateau}. The frequency
distribution $\tilde{\omega}_{n}$ (a) shows a small plateau of
constant frequency over about 60 oscillators towards the high frequency end of
the distribution; these are the frequency locked oscillators. The
synchronization index (b) shows that $\chi_{n}$ approaches unity for
most of the locked oscillators, which means that $z_{n}$ for these oscillators
is essentially time independent once the rotation of the phase at the order
parameter frequency is subtracted out. A careful scrutiny of the two panels
reveals that the plateau in $\tilde{\omega}_{n}$ is sharper and more extended
than the one in $\chi_{n}$, so that not all the frequency locked oscillators
are time independent in the rotating frame. The dynamical state is actually
quite complicated, as a review of $z_{n}(t)$ shows.%

\begin{figure}[tbh]
\begin{center}
\includegraphics[width=3.0in]{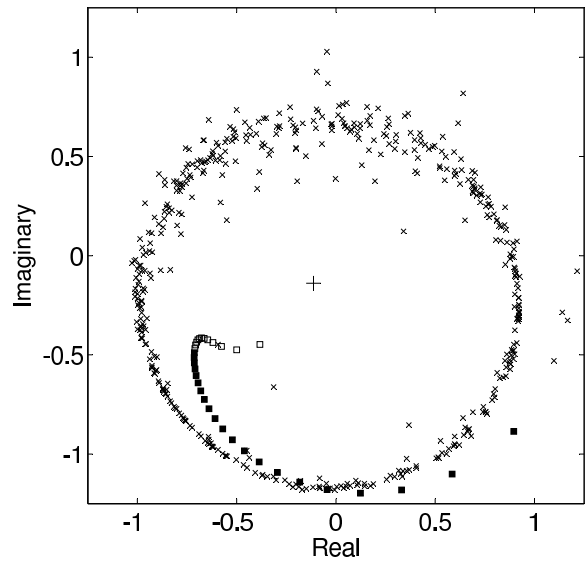}%
\caption{Snapshot of $z_{n}$ and order parameter $\Psi$ in the complex plane
at a time after transients have decayed. The \textquotedblleft%
+\textquotedblright\ towards the center of the plot gives the
value of the order parameter. The squares are oscillators locked
to the order parameter frequency (i.e.\ those with
$\tilde{\omega}_{n}=\Omega$): solid squares are stationary when
the $e^{i{\Omega}t}$ dependence is removed, open squares show an
additional dynamics rotating around the tail of the fixed
distribution in the rotating system. The \textquotedblleft$\times
$\textquotedblright\ are oscillators that are not locked to the
order
parameter.}%
\label{l_matlab_fig2}%
\end{center}
\end{figure}
Figure \ref{l_matlab_fig2} shows a plot of a snapshot of the
complex amplitude $z_{n}$ of each oscillator. The
\textquotedblleft+\textquotedblright\ is the order parameter. The
other points correspond to $z_{n}$ for each oscillator. It is
useful to study the dynamics of this plot after rotation at the
mean order parameter frequency is eliminated. As time evolves, the
square symbols remain fixed in such a plot (except for very small
fluctuations): these represent oscillators that are locked to the
order parameter. For the solid squares the complex amplitudes are
essentially time independent once the phase of the order parameter
is extracted. The oscillators represented by an
\textquotedblleft$\times$\textquotedblright\ on the other hand
rotate clockwise or anticlockwise about the origin: these
correspond to unlocked or running oscillators. The open squares
exhibit a more complicated dynamics undergoing small amplitude
orbits around the tail of the locked oscillator distribution.
These oscillators are locked to the order parameter, since the
difference of their phases from the order parameter phase does not
drift over arbitrary long times. The values of
$\tilde{\omega}_{n}$ for these oscillators are on the locked
plateau in Fig.~\ref{Fig_smr_plateau}. However the amplitudes are
not constant in the rotating plot, the values of $\chi_{n}$ are
less than unity, and the oscillators contribute to fluctuations of
the order parameter. Thus the fluctuations of $R(t)$ shown in
Fig.\ \ref{Fig_Orderparameters} are not just due to finite $N$
effects, and we believe they would persist in the
$N\rightarrow\infty$ limit. We have not explored larger $N$ to pin
down whether these intrinsic fluctuations are periodic or
aperiodic in the large $N$ limit.

As we discuss in more detail later, for a bounded distribution of
frequencies it is possible to find a low amplitude synchronized
state with $R\neq0$, but with a smooth frequency distribution
showing that there is no frequency locking. In fact for this state
the distribution of actual oscillator frequencies does not overlap
the order parameter frequency. The nonzero order parameter is
caused by a systematic slowing of the phase rotation of the
oscillators in the vicinity of the order parameter phase, rather
than by a fraction of the phases becoming locked to the order
parameter phase. For an unbounded distribution such as the
Lorentzian, this same effect occurs but is supplemented by the
small fraction of locked oscillators.
\begin{figure}[tbh]
\begin{center}
\includegraphics[width=3in]{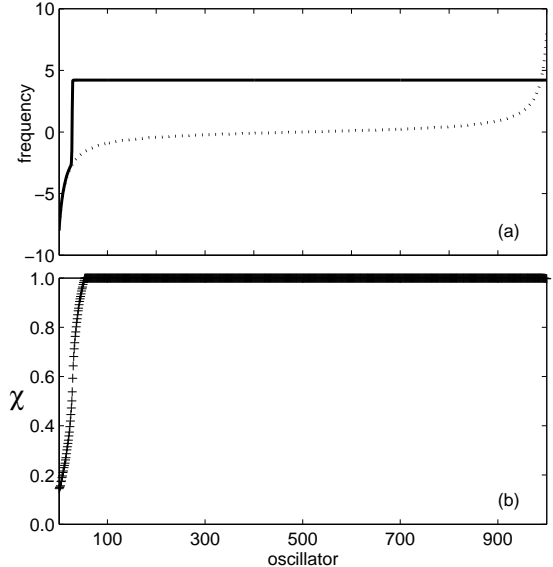}%
\caption{Same as in Fig.~\ref{Fig_smr_plateau} but for the large amplitude
synchronized state.}%
\label{Fig_lgr_plateau}%
\end{center}
\end{figure}

For the same parameter values as for the previously described low
amplitude synchronized state for the Lorentzian distribution there
is a second dynamical state that may be reached depending on the
initial conditions of the simulation. This state has a much larger
value of the order parameter $\left\langle R\right\rangle
_{t}\simeq0.9$ and the plot of the distribution of frequencies,
Fig.~\ref{Fig_lgr_plateau}(a), shows a correspondingly larger
fraction of locked oscillators. (This should be compared with
panel (a) of Fig.~\ref{Fig_smr_plateau}.) For the cutoff
Lorentzian distribution used, all the oscillators at the high
frequency end of the distribution are locked, leaving only a small
fraction of free running oscillators at the low end of the
frequency distribution. For the $N\rightarrow\infty$ limit and a
Lorentzian distribution without a cutoff, unlocked oscillators
remain at both the high and low frequency ends of the
distribution. For a bounded distribution of bare frequencies, such
as triangular or top-hat, a fully locked state in which all the
oscillators rotate with the same frequency and fixed magnitude may
be found.

\section{Formulation of Synchronization}

\label{sec_form} We now turn to the analysis of the synchronization of
oscillators described by Eq.~(\ref{eq_cross}). Since we are interested in the
behavior for a large number of oscillators, it is convenient to go to a
continuum description, where we label the oscillators by their uncoupled
linear frequency $\omega=\omega_{n}$ rather than the index $n$, $z_{n}%
\rightarrow z(\omega)$. Introducing the order parameter
Eq.~(\ref{OrderParameter}), the oscillator equations can be written in
magnitude-phase form as
\begin{subequations}
\label{theta_r}%
\begin{align}
d_{t}\bar{\theta}  &  =\bar{\omega}+\alpha(1-r^{2})+\frac{\beta R}{r}\cos
\bar{\theta}\label{theta}\\
d_{t}r  &  =(1-r^{2})r+\beta R\sin\bar{\theta} \label{r}%
\end{align}
where $\bar{\theta}=\theta-\Theta$ is the oscillator phase relative to that of
the order parameter as before, and $\bar{\omega}$ is the bare oscillator
frequency shifted by $\alpha+\beta$ and measured relative to the order
parameter frequency $\Omega=\dot{\Theta}$
\end{subequations}
\begin{equation}
\bar{\omega}=\omega-\alpha-\beta-\Omega.
\end{equation}
Note that if the order parameter is zero $R=0$, the magnitude $r$
will relax to $1$, and the $n$th oscillator will evolve at the
frequency $\omega _{n}-\alpha-\beta$. The first component of this
frequency shift is just the nonlinear shift at $r=1$, and the
second is from the interaction of each oscillator with the
incoherent motion of the other oscillators.

At each time $t$ the oscillators are specified by the distribution
$\rho(r,\bar{\theta},\bar{\omega},t)$ where $\rho(r,\bar{\theta},\bar{\omega
},t)r\,dr\,d\bar{\theta}$ is the fraction of the oscillators with shifted
frequency $\bar{\omega}$ that at time $t$ have magnitude between $r$ and
$r+dr$ and shifted phase between $\bar{\theta}$ and $\bar{\theta}+d\bar
{\theta}$. The order parameter is given by the self-consistency condition%
\begin{equation}
R\,=\left\langle re^{i\bar{\theta}}\right\rangle =\int d\bar{\omega}\bar
{g}(\bar{\omega})\int r\,dr\,d\bar{\theta}\rho(r,\bar{\theta},\bar{\omega
},t)re^{i\bar{\theta}}.\, \label{SelfConsistency}%
\end{equation}
where $\bar{g}(\bar{\omega})$ is the distribution of oscillator frequencies
expressed in terms of the shifted frequency $\bar{\omega}$. It is useful to
split this expression into real and imaginary parts. The imaginary part is%
\begin{equation}
\int d\bar{\omega}\bar{g}(\bar{\omega})\int r\,dr\,d\bar{\theta}\rho
(r,\bar{\theta},\bar{\omega},t)r\sin\bar{\theta}=0. \label{ImSelfConsistency}%
\end{equation}
Because the phases $\bar\theta$ are measured relative to the
orientation of the order parameter, this expresses the fact that
the components of the complex amplitudes $r\sin\bar\theta$ normal
to the order parameter must average to zero. Note that unlike the
cases of the Kuramoto model [Eq.~(\ref{Kuramoto})] and the
dissipatively coupled complex amplitude model
[Eq.~(\ref{AmplitudePhase})] studied by Matthews et al., this
condition is not trivially satisfied even for the case of a
symmetric distribution $g(\omega)$, and in fact serves to
determine the frequency $\Omega$ of the order parameter. The
frequency $\Omega$ is also not trivially related to the mean
frequency of
the oscillator distribution. The real part of Eq.~(\ref{SelfConsistency}) is%
\begin{equation}
\int d\bar{\omega}\bar{g}(\bar{\omega})\int r\,dr\,d\bar{\theta}\rho
(r,\bar{\theta},\bar{\omega},t)r\cos\bar{\theta}=R. \label{ReSelfConsistency}%
\end{equation}
This is the condition that the components $r\cos\bar\theta$ along
the direction of the order parameter must average to the magnitude
$R$. This condition serves to self-consistently fix the value of
$R$.

The expectation that Eqs.~(\ref{theta_r}) might lead to
synchronization follows from the behavior for narrow frequency
distributions and large $\alpha$. If the width of the distribution
of frequencies is small compared to the relaxation rate of the
magnitude, which is of order one in the time units used in
Eq.~(\ref{eq_cross}), the magnitude relaxes rapidly to the value
given
by the instantaneous value of $\bar{\theta}$, i.e.~to the solution of%
\begin{equation}
(1-r^{2})r=-\beta R\sin\bar{\theta}. \label{rConstant}%
\end{equation}
If $r$ is close to one, which we will see applies at the onset of
synchronization for large $\alpha$, this gives%
\begin{equation}
1-r^{2}\simeq-\beta R\sin\bar{\theta}.
\end{equation}
In this case Eq.~(\ref{theta}) becomes, ignoring $\beta R$ compared with
$\alpha\beta R$,%
\begin{equation}
d_{t}\bar{\theta}\simeq\bar{\omega}-\alpha\beta R\sin\bar{\theta}.
\label{ApproxKuramoto}%
\end{equation}
This equation is the same as the one derived from the Kuramoto
model in Eq.~(\ref{KuramotoField}) with $\alpha\beta$ playing the
role of the coupling constant $K$, and therefore predicts an onset
of synchronization at $\alpha\beta=2(\pi g(0))^{-1}$ \cite{K75}.

To uncover more fully the behavior of Eq.~(\ref{eq_cross}) we consider two
issues: the onset of synchronization, detected as the linear instability of
the unsynchronized $R=0$ state, as a function of $\alpha$ and $\beta$ for some
given frequency distribution $g(\omega)$; and the existence of a fully locked
state for large values of $\alpha\beta$.

\section{Onset of synchronization}

\label{sec_onset} We first consider the initial onset of partial
synchronization from the unsynchronized state in which each
oscillator runs at its own frequency $\omega_{n}$. We identify
synchronization through a nonzero value of the order parameter.
This often arises because a finite fraction of the oscillators
become locked to the same frequency, which we call a partially
(fully if the fraction is unity) locked state. However we also
find situations where $R\neq0$, but there is no frequency locking.

The onset of synchronization can be determined by a linear instability
analysis of the unsynchronized state. This is calculated by linearizing the
distribution $\rho$ around the unsynchronized distribution, which is a uniform
phase distribution at $r=1$, and seeking the parameter values at which
deviations from the uniform phase distribution begin to grow exponentially.
This follows the method of Matthews et al. \cite{MMS91}, although care is
needed in the analysis due to the more important role the magnitude
perturbations $r$ play in the present case.

Introducing the small expansion parameter $\varepsilon$ characterizing the
small deviations from the unsynchronized state, we write%
\begin{equation}
\rho(r,\theta,\bar{\omega},t)=(2\pi r)^{-1}\delta\lbrack r-1-\varepsilon
r_{1}(\bar{\theta},\bar{\omega},t)][1+\varepsilon f_{1}(\bar{\theta}%
,\bar{\omega},t)]. \label{RhoExpansion}%
\end{equation}
Note that for $\varepsilon=0$ this does indeed give the appropriately
normalized distribution for the unsynchronized state, in which all the
oscillators have unit magnitude, $r=1$, and the phase distribution is
constant. Also, for nonzero $\varepsilon$, $\rho$ remains normalized to linear
order%
\begin{equation}
\int_{0}^{\infty}dr\int_{0}^{2\pi}d\theta\,r\,\rho(r,\theta,\bar{\omega
},t)=1+O(\varepsilon^{2}),
\end{equation}
providing the average of $f_{1}$ over $\bar{\theta}$ is zero.

The equation for the evolution of the radial perturbation of the distribution
$r_{1}$ is given by noting that%
\begin{equation}
\frac{dr}{dt}=\frac{\partial r}{\partial\bar{\theta}}d_{t}\bar{\theta}%
+\frac{\partial r}{\partial t}.
\end{equation}
The left hand side is evaluated by expanding Eq.~(\ref{r}) to first order in
$\varepsilon$ and also expanding the magnitude of the order parameter
$R=\varepsilon R_{1}+\cdots$. , and the right hand side by the replacement
$d_{t}\bar{\theta}=\bar{\omega}+O(\varepsilon)$ and assuming an exponential
growth or decay of the perturbation $\partial r/\partial t=\varepsilon\partial
r_{1}/\partial t=\lambda\varepsilon r_{1}$. The result is%
\begin{equation}
\frac{\partial r_{1}}{\partial\bar{\theta}}\bar{\omega}+(\lambda+2)r_{1}=\beta
R_{1}\sin\bar{\theta}. \label{r1}%
\end{equation}
Equation (\ref{r1}) is solved by%
\begin{equation}
r_{1}=R_{1}(A\cos\bar{\theta}+B\sin\bar{\theta}), \label{r1AB}%
\end{equation}
with
\begin{subequations}
\label{AB}%
\begin{align}
A  &  =-\beta\frac{\bar{\omega}}{\bar{\omega}^{2}+(\lambda+2)^{2}},\label{A}\\
B  &  =\beta\frac{(\lambda+2)}{\bar{\omega}^{2}+(\lambda+2)^{2}}. \label{B}%
\end{align}

To extract the equation for $f_{1}$ integrate the equation for the
conservation of probability%
\end{subequations}
\begin{equation}
\frac{\partial\rho}{\partial t}+\mathbf{\nabla}\cdot(\rho\mathbf{v})=0
\label{RhoConservation}%
\end{equation}
over radius. Here $\mathbf{v}$ is the velocity in complex
amplitude space, which in polar coordinates is
$(d_{t}r,rd_{t}\bar{\theta})$ given by Eqs.~(\ref{theta_r}). Again
replacing $\partial f_{1}/\partial t$ by $\lambda f_{1}$, and
evaluating $r_{1}$ from Eqs.~(\ref{r1AB},\ref{AB}), this
gives at $O(\varepsilon)$%
\begin{equation}
\lambda f_{1}+\bar{\omega}\partial_{\bar{\theta}}f_{1}=2\alpha R_{1}%
(-A\sin\bar{\theta}+B\cos\bar{\theta})+\beta R_{1}\sin\bar{\theta}.
\end{equation}
This is solved by%
\begin{equation}
f_{1}=R_{1}(C\cos\bar{\theta}+D\sin\bar{\theta})
\end{equation}
with
\begin{subequations}
\label{CD}%
\begin{align}
C  &  =\beta\frac{2\alpha(\lambda^{2}+2\lambda-\bar{\omega}^{2})-\bar{\omega
}[\bar{\omega}^{2}+(\lambda+2)^{2}]}{\left(  \bar{\omega}^{2}+\lambda
^{2}\right)  \left[  \bar{\omega}^{2}+(\lambda+2)^{2}\right]  },\label{C}\\
D  &  =\beta\frac{4\alpha\bar{\omega}(\lambda+1)+\lambda\lbrack\bar{\omega
}^{2}+(\lambda+2)^{2}]}{\left(  \bar{\omega}^{2}+\lambda^{2}\right)  \left[
\bar{\omega}^{2}+(\lambda+2)^{2}\right]  }. \label{D}%
\end{align}

We now evaluate the self-consistency condition Eq.~(\ref{SelfConsistency}) to
first order in $\varepsilon$. The imaginary part is%
\end{subequations}
\begin{equation}
\int
d\bar{\omega}\,\bar{g}(\bar{\omega})\int_{0}^{2\pi}d\bar{\theta
}\,(1+\varepsilon r_{1}+\cdots)(1+\varepsilon f_{1}+\cdots
)\sin\bar{\theta}=0, \label{Perp}%
\end{equation}
which to first order in $\varepsilon$ gives%
\begin{equation}
\int d\bar{\omega}\,\bar{g}(\bar{\omega})\left[  B+D\right]  =0.
\label{PerpInt}%
\end{equation}
Similarly the real part of the self-consistency condition is%
\begin{equation}
\int d\bar{\omega}\,\bar{g}(\bar{\omega})\left[  A+C\right]  =2.
\label{ParInt}%
\end{equation}

We want to evaluate Eqs.~(\ref{PerpInt}) and (\ref{ParInt}) at the onset of
instability where the growth rate $\lambda\rightarrow0$. We can set
$\lambda=0$ in Eqs.~(\ref{PerpInt},\ref{ParInt}) with Eqs.~(\ref{AB}) and
(\ref{CD}) except in terms with $\bar{\omega}^{2}+\lambda^{2}$ in the
denominator, since such terms may give large contributions to the integral
from the region of small $\bar{\omega}$. A term involving just $\lambda
/(\bar{\omega}^{2}+\lambda^{2})$ gives a finite integral, but if this is
multiplied by powers of $\bar{\omega}$ or $\lambda$ the integral goes to zero
in the $\lambda\rightarrow0$ limit. Similarly for a term involving
$\bar{\omega}/(\bar{\omega}^{2}+\lambda^{2})$ we must take the limit
$\lambda\rightarrow0$ after doing the integral (this is equivalent to the
principal value integral), whereas if this term is multiplied by powers of
$\bar{\omega}$ we can put $\lambda=0$ immediately.

The needed integrals are
\begin{subequations}
\label{integrals}%
\begin{align}
I_{1}  &  =\int\bar{g}(\bar{\omega})\frac{1}{\bar{\omega}^{2}+4},\\
I_{2}  &  =\lim_{\lambda\rightarrow0}\int\bar{g}(\bar{\omega})\frac
{\bar{\omega}}{\bar{\omega}^{2}+\lambda^{2}},\\
I_{3}  &  =\int\bar{g}(\bar{\omega})\frac{\bar{\omega}}{\bar{\omega}^{2}+4},\\
I_{4}  &  =\lim_{\lambda\rightarrow0}\int\bar{g}(\bar{\omega})\frac{\lambda
}{\bar{\omega}^{2}+\lambda^{2}}=\pi\bar{g}(\bar{\omega}=0).
\end{align}

The imaginary part of the self consistency condition
Eq.~(\ref{PerpInt})
becomes%
\end{subequations}
\begin{equation}
2I_{1}+\alpha I_{2}-\alpha I_{3}+I_{4}=0, \label{perp_integrals}%
\end{equation}
and the real part reduces to the condition for $\beta_{c}$%
\begin{equation}
\beta_{c}=2(-I_{3}+\alpha I_{4}-2\alpha I_{1}-I_{2})^{-1}.
\label{par_integrals}%
\end{equation}

We have explicitly evaluated the integrals for top-hat, triangular, and
Lorentzian distributions of bare frequencies. These results will be presented
after we discuss full locking.

\section{Full Locking}

\label{Sec_FullLocking} We define the fully locked state as one in
which all the phases are rotating at the same frequency as the
order parameter, and the magnitudes are constant in time. These
solutions are defined by Eq.~(\ref{theta}) with $d_t\bar\theta=0$,
which with Eq.~(\ref{rConstant}) can be written%
\begin{equation}
\bar{\omega}=\frac{\beta R}{r}(\alpha\sin\bar{\theta}-\cos\bar{\theta}%
)=F(\bar{\theta}), \label{Locked}%
\end{equation}
where the solution to the cubic equation (\ref{rConstant}) for
$r(\bar\theta)$ is to be used to form the function of phase alone
$F(\bar{\theta})$. The function $F(\bar{\theta})$ acts as the
force pinning the locked oscillators to the order parameter,
generalizing the notion introduced below
Eq.~(\ref{KuramotoField}), and plays a central role in our
discussion of locking. A particular oscillator, identified by its
shifted frequency $\bar{\omega}$, will be locked to the order
parameter if Eq.~(\ref{Locked}) has a solution
$\bar{\theta}=F^{-1}(\bar{\omega})$ [and then $r$ is given by
solving Eq.~(\ref{rConstant})] and if this solution is stable. The
stability is tested by linearizing Eqs.~(\ref{theta_r}) about the
solution. The fully locked solution will only exist if stable,
locked solutions to Eq.~(\ref{Locked}) exist for \emph{all} the
oscillators in the distribution. We are thus led to investigate
the properties of the function $F(\bar{\theta })$. In addition the
self consistency condition Eq.~(\ref{SelfConsistency}) must be
satisfied.

For small $\beta R$, the magnitude $r(\bar{\theta})$ given by
Eq.~(\ref{rConstant}) remains bounded away from zero for all $\bar{\theta}$,
and the function $F(\bar{\theta})$ varies continuously between minimum and
maximum values $F_{\min}\leq F\leq F_{\max}$. In this case, we immediately see
that the fully locked solution only occurs for bounded distributions, $\bar
{g}(\bar{\omega})$ nonzero only between finite $\bar{\omega}_{\min}$ and
$\bar{\omega}_{\max}$. In such cases we define $\bar{\omega}_{\max}%
-\bar{\omega}_{\min}$, which is equal to the range of unshifted frequencies
$\omega_{\max}-\omega_{\min}$, as $w$ the width of the distribution. More
generally, although $F(\bar{\theta})$ can vary over an infinite range (because
$r$ may become zero) we find that only a finite range yields stable solutions,
so that again complete locking only occurs for a bounded distribution of
oscillator frequencies.

We first look at the fully locked solution for large values of
$\alpha\beta$. In this case the phases of the locked oscillators
cover a narrow range of angles, since the range of the pinning
force $F$ becomes large for large $\alpha\beta$. The imaginary
part of the self consistency condition
Eq.~(\ref{ImSelfConsistency}) shows that the range of phases must
be around $\bar{\theta}=0$. Equation (\ref{Locked}) can now be
approximated by expanding
around $\bar{\theta}=0$ (note $r\simeq1$ here) and becomes for large $\alpha$%
\begin{equation}
\bar{\omega}=\omega-\alpha-\beta-\Omega\simeq-\beta R(1-\alpha\bar{\theta}).
\label{AlphaLarge}%
\end{equation}
The imaginary part of the self-consistency condition reduces to $\left\langle
\bar{\theta}\right\rangle =0$ (the average is over the distribution of
frequencies), and the real part to simply $R\simeq1$. Finally, averaging
Eq.~(\ref{AlphaLarge}) over the distribution of frequencies fixes the order
parameter frequency (the common frequency of all the oscillators)%
\begin{equation}
\Omega\simeq\left\langle \omega\right\rangle -\alpha.
\end{equation}
Thus the order parameter, and all the oscillators, evolve at a frequency given
by the mean of the distribution $g(\omega)$ shifted by the nonlinear effect
for $r=1$.

We now investigate the limit to the fully locked regime as we lower $\alpha$.
We first summarize the argument, and then present the details. The fully
locked solutions are determined by a rather complicated set of interconnected
equations. They can be found by the following algorithm. For fixed values of
$\alpha,B=\beta R$ solve Eq.~(\ref{rConstant}) for real positive
$r(\bar{\theta})$ and hence calculate $F(\bar{\theta})$. Using
Eqs.~(\ref{theta_r}) the stability of each solution is tested: the eigenvalues
of this analysis are (see Eq.~(\ref{Eigenvalues_2}) below)%
\begin{equation}
\lambda_{\pm}=1-2r^{2}\pm\sqrt{1-\frac{B^{2}}{r^{2}}-2r^{2}+2r^{4}-2\alpha
rB\cos\bar{\theta}}, \label{Eigenvalues}%
\end{equation}
with $\cos\bar{\theta}=\pm\sqrt{1-r^{2}(1-r^{2})^{2}/B^{2}}$. From this we can
identify a range $F_{\min}^{(s)}<F<F_{\max}^{(s)}$ corresponding to the range
of existence and stability of locked oscillators. A fully locked solution must
then satisfy the constraints%
\begin{equation}
\bar{\omega}_{\min}\geq F_{\min}^{(s)};~\bar{\omega}_{\max}\leq F_{\max}%
^{(s)}, \label{OmegaCondition}%
\end{equation}
together with the condition given by the imaginary part of the self
consistency condition%
\begin{equation}
\int_{\bar{\omega}_{\min}}^{\bar{\omega}_{\max}}\bar{g}(\bar{\omega})\,r%
(\bar{\omega})\sin[\bar{\theta}(\bar{\omega})]\,d\bar{\omega}=0,
\label{Transverse}%
\end{equation}
where $\bar{\theta}(\bar{\omega})=F^{-1}(\bar{\omega})$, and
$r(\bar{\omega})$ is then the solution to Eq.~(\ref{rConstant}).
The boundary of the fully locked region occurs either when
$\bar{\omega}_{\min}=F_{\min}^{(s)}$ or when
$\bar{\omega}_{\max}=F_{\max}^{(s)}$. This equation can be
interpreted as fixing the order parameter frequency $\Omega$ in
terms of $\alpha$ and $B$. Notice that Eq.~(\ref{Eigenvalues})
shows that for $r<1/\sqrt{2}$ one eigenvalue certainly has a
positive real part indicating instability, so that for stable
solutions $r\geq1/\sqrt{2}$ and $F_{\min}^{(s)}$ is finite. Since
$\bar{\omega }_{\min}$ or $\bar{\omega}_{\max}$ is now determined,
and $\bar{\omega}_{\max }-\bar{\omega}_{\min}=w$,
Eq.~(\ref{Transverse}) is an implicit equation relating the values
of $\alpha$, $B$ and $w$ at the locking transition. To
complete the solution, the real part of the self consistency equation%
\begin{equation}
\int_{\bar{\omega}_{\min}}^{\bar{\omega}_{\max}}\bar{g}(\bar{\omega})\,r%
(\bar{\omega})\cos[\bar{\theta}(\bar{\omega})]\,d\bar{\omega}=R
\label{Longitudingal}%
\end{equation}
then serves to fix $R$ at locking, from which the value of $\beta=B/R$ at the
transition to full locking can be found.

\subsection{Existence of individual oscillator locked solution}

We first consider the \emph{existence} of a locked solution for an
individual oscillator, i.e.\ a \emph{stationary} solution of
Eqs.~(\ref{theta_r}). Equation (\ref{rConstant}) gives the cubic
equation for $r(\bar{\theta})$ for
each $B=\beta R$%
\begin{equation}
(1-r^{2})r+B\sin\bar{\theta}=0. \label{r_theta}%
\end{equation}
Of course, for the physical solution $r$ must be real and positive. Thus we
need to analyze the properties of the real positive solutions to%
\begin{equation}
(1-r^{2})r+X=0, \label{cubic}%
\end{equation}
as $X$ varies.%

\begin{figure}
[tbh]
\begin{center}
\includegraphics[width=2.8in]{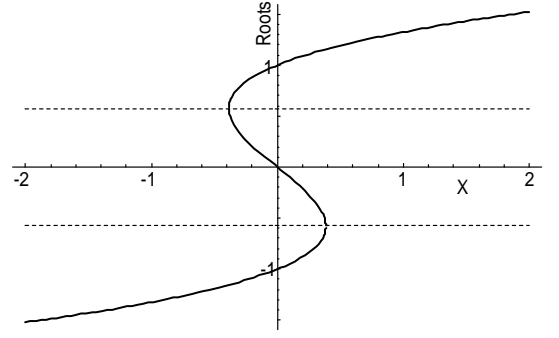}%
\caption{Solutions to the cubic equation Eq.~(\ref{cubic}) as $X$ varies. The
dashed lines are $y=\pm1/\sqrt{3}$.}%
\label{cubic_fig}%
\end{center}
\end{figure}
For $X=0$ the solutions to the cubic are $r=\pm1,0$. As $\left\vert
X\right\vert $ increases, the solutions remain real until two of the roots
collide and become complex. Since the sum of the roots to Eq.~(\ref{cubic}) is
zero, at the collision the equation takes the form%
\begin{equation}
(r-a)^{2}(r+2a)=0.
\end{equation}
Matching coefficients shows that collision occurs at $X=\pm\frac{2}{\sqrt{27}%
}$ when $r=\mp\frac{1}{\sqrt{3}}$. The form of Eq.~(\ref{cubic}) is actually
already in what is known as the \textquotedblleft
depressed\ form\textquotedblright\ of a cubic equation, for which the solution
is relatively simple. Inspecting the form of these solutions shows that for
$\left\vert X\right\vert >\frac{2}{\sqrt{27}}$ there is one real solution and
a complex pair in the form $2a,-a\pm ib$. The product of the roots is
determined by the constant in the cubic, giving%
\begin{equation}
2a(a^{2}+b^{2})=X
\end{equation}
and so for $X<-\frac{2}{\sqrt{27}}$ the real root is negative, and for
$X>\frac{2}{\sqrt{27}}$ the real root is positive. These results are confirmed
by numerical solution as shown in Fig.~\ref{cubic_fig}.%

\begin{figure}
[tbh]
\begin{center}
\includegraphics[width=2.8in]{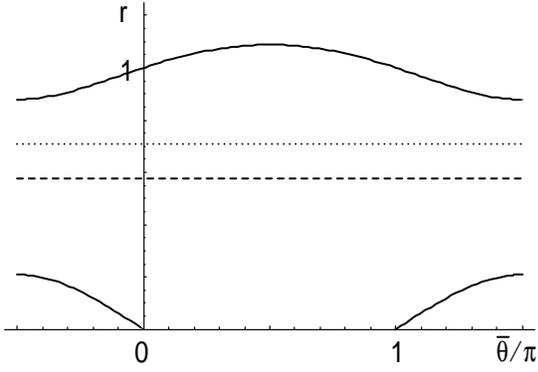}%
\caption{Plot of $r(\bar{\theta})$ for $B<2/\sqrt{27}$. The dashed line is
$r=1/\sqrt{3}$ and the dotted line $r=1/\sqrt{2}$. The condition $r>1/\sqrt
{2}$ is necessary (but not sufficient) for the solution to be stable. The
lower branch always satisfies $r<1/\sqrt{3}$, and so only the upper branch may
have stable regions.}%
\label{r_theta_Bs}%
\end{center}
\end{figure}
\begin{figure}
[tbh]
\begin{center}
\includegraphics[width=2.8in]{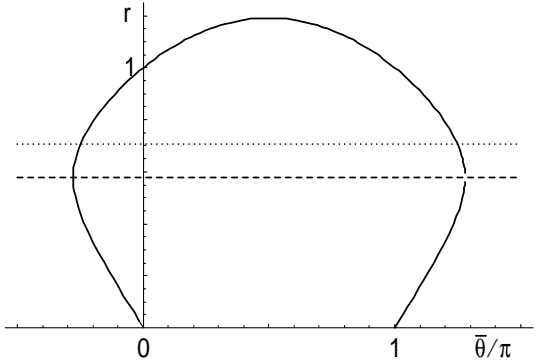}%
\caption{Plot of $r(\bar{\theta})$ for $B>2/\sqrt{27}$. Solutions exist for
$-\theta_{B}\leq\bar{\theta}\leq\pi+\theta_{B}$ with $\theta_{B}=\sin
^{-1}(2/B\sqrt{27})$. The dashed line is $r=1/\sqrt{3}$ and the dotted line
$r=1/\sqrt{2}$.}%
\label{r_theta_Bl}%
\end{center}
\end{figure}
Thus we find the following behavior for the real positive
solutions to Eq.~(\ref{r_theta}). For $B<2/\sqrt{27}$ there is a
root that is positive with $r>1/\sqrt{3}$ for all $\bar{\theta}$,
a root that varies between positive and negative values with
magnitude less than $1/\sqrt{3}$, and a root that is negative for
all $\bar{\theta}$. We have seen that the stability analysis shows
that any solution with $r<1/\sqrt{2}$ is unstable$.$ Thus only the
first root is relevant (see Fig.\ \ref{r_theta_Bs}). For
$B>2/\sqrt{27}$ real positive roots exist for
$-\theta_{B}\leq\bar{\theta}\leq\pi+\theta_{B}$ with
$\theta_{B}=\sin^{-1}\frac{2}{B\sqrt{27}}$. In the range where
there are two real positive roots,
$-\theta_{B}\leq\bar{\theta}\leq0$ and $\pi\leq
\bar{\theta}\leq\pi+\theta_{B}$, only the larger may be in the
stable range $r>1/\sqrt{2}$, (see Fig.\ \ref{r_theta_Bl}).

Having found the stationary solutions $r(\bar{\theta})$, the condition for
stationary phase $d_{t}\bar{\theta}=0$ can be written%
\begin{equation}
\bar{\omega}=F(\bar{\theta}) \label{omega_F}%
\end{equation}
with%
\begin{equation}
F(\bar{\theta})=\frac{\beta R}{r(\bar{\theta})}(\alpha\sin\bar{\theta}%
-\cos\bar{\theta}) \label{F_bar_theta}.
\end{equation}
The function $F(\bar{\theta})$ acts as the force acting on each oscillator
tending to pin its frequency to the order parameter frequency. The range of
possible $\bar{\omega}$ for locked oscillators is limited by the range of $F$
corresponding to stable stationary solutions.

\subsection{Stability of individual oscillator locked solution}

The stability of the locked solution for an individual oscillator
mentioned in the previous section is given by linearizing
Eqs.~(\ref{theta_r}) about the
stationary solutions. The linearized equations are%
\begin{align}
d_{t}\delta\bar{\theta}  &  =-\left(  \frac{\beta
R}{r}\sin\bar{\theta }\right)  \,\delta\bar{\theta}-\left(
2\alpha r+\frac{\beta R}{r^{2}}\cos
\bar{\theta}\right)  \delta r\\
d_{t}\delta r  &  =(\beta R\cos\bar{\theta})\,\delta
\bar{\theta}+(1-3r^{2})\delta r%
\end{align}
The eigenvalues are%
\begin{equation}
\lambda_{\pm}=1-2r^{2}\pm\sqrt{S}\label{Eigenvalues_2}%
\end{equation}
with
\begin{equation}
S=1-\frac{B^{2}%
}{r^{2}}-2r^{2}+2r^{4}-2\alpha Br\cos\bar{\theta}
\end{equation}
This immediately shows us $r>1/\sqrt{2}$ is a necessary condition for
stability, since $\operatorname{Re}\lambda_{+}\geq1-2r^{2}$ (with the equality
if $S$ is negative so that $\lambda_{\pm}$ are complex).

Let us first seek the condition for a root $\lambda_{\pm}$ to become zero,
signaling a stationary bifurcation. It is convenient to go back to the
original equations Eqs.~(\ref{theta_r}) in the form%
\begin{subequations}
\label{linear}
\begin{align}
d_{t}\bar{\theta}  &  =\bar{\omega}-f(r,\bar{\theta})\\
d_{t}r  &  =(1-r^{2})r+\beta R\sin\bar{\theta}%
\end{align}
\end{subequations}
with%
\begin{equation}
f(r,\bar{\theta})=-\alpha(1-r^{2})-\frac{\beta R}{r}\cos\bar{\theta}.
\end{equation}
Then the determinant of the linear matrix derived from Eqs.~(\ref{linear}) is%
\begin{equation}
D=\left\vert
\begin{array}
[c]{cc}%
-\frac{\partial f}{\partial\bar{\theta}} & -\frac{\partial f}{\partial r}\\
\beta R\cos\bar{\theta} & 1-3r^{2}%
\end{array}
\right\vert =-(1-3r^{2})\frac{\partial f}{\partial\bar{\theta}}+\beta
R\cos\bar{\theta}\frac{\partial f}{\partial r}. \label{det}%
\end{equation}
The stationary solution $r(\bar{\theta})$ satisfies Eq.~(\ref{r_theta}) and so
Eq.~(\ref{det}) can be written%
\begin{equation}
D=-(1-3r^{2})\frac{dF}{d\bar{\theta}},
\end{equation}
since%
\begin{equation}
\frac{dF(\bar{\theta})}{d\bar{\theta}}=\frac{df(\bar{\theta},r(\bar{\theta}%
))}{d\bar{\theta}}=\frac{\partial f}{\partial\bar{\theta}}+\frac{\partial
f}{\partial r}\frac{dr}{d\bar{\theta}}.
\end{equation}
Thus a zero eigenvalue occurs at and only at stationary points of
$F(\bar{\theta})$ or at $r=1/\sqrt{3}$. The latter is where $r(\bar{\theta})$
has a vertical tangent (cf.\ Fig.\ \ref{r_theta_Bl}) but always occurs outside
the range of stable solutions, for which we know $r>1/\sqrt{2}$.

The only other possibility for an instability is $\operatorname{Re}%
\lambda_{\pm}=0$, $\operatorname{Im}\lambda_{\pm}\neq0$. This can
occur only at $r=1/\sqrt {2}$, and if $S<0$.

Another result can be derived: if $dF/d\bar{\theta}<0$ and $r>1/\sqrt{3}$ then
the determinant $D<0$. This implies that the eigenvalues are real (since the
product of a complex conjugate pair is always positive), one positive and one
negative. Thus a negative slope of $F(\bar{\theta})$ implies instability. Also
the Hopf bifurcation can only occur at values of $\bar{\theta}$ where
$dF/d\bar{\theta}>0$.

To satisfy the imaginary part of the self-consistency condition
Eq.~(\ref{Transverse}), the range of phases of locked oscillator phases must
straddle $\bar{\theta}=0$. The oscillator solution here, $\bar{\theta}=0,r=1$
is always stable for positive $\alpha,B$, since here $\lambda_{+}%
=-1+\sqrt{1-B(B+2\alpha)}$. Thus the range of possible stable stationary
solutions for locked oscillators is given by the range of $\bar{\theta}$
bounded on either side of $\bar{\theta}=0$ by the closest stationary
bifurcation point or by a Hopf bifurcation occurring at $r=1/\sqrt{2}$,
whichever is closest.

\subsection{Properties of the locking force}

We now derive the properties of the locking force, $F(\bar{\theta})$. For
small $B=\beta R$, Eq.~(\ref{rConstant}) yields a stationary solution with
$r(\bar{\theta})\simeq1$ and so%
\begin{align}
F(\bar{\theta})  &  \simeq B(\alpha\sin\bar{\theta}-\cos\bar{\theta})\\
&  =B\sqrt{\alpha^{2}+1}\sin(\bar{\theta}-\theta_{\alpha})
\end{align}
with $\theta_{a}=\tan^{-1}(1/\alpha)$. Thus $F(\bar{\theta})$ is a sinusoidal
function of angle in this limit. The stable solutions are the positive slope
region for $\bar{\theta}$ between $\theta_{\alpha}-\pi/2$ and $\theta_{\alpha
}+\pi/2$.%

\begin{figure*}
[tbh]
\begin{center}
\includegraphics[width=5.0in]{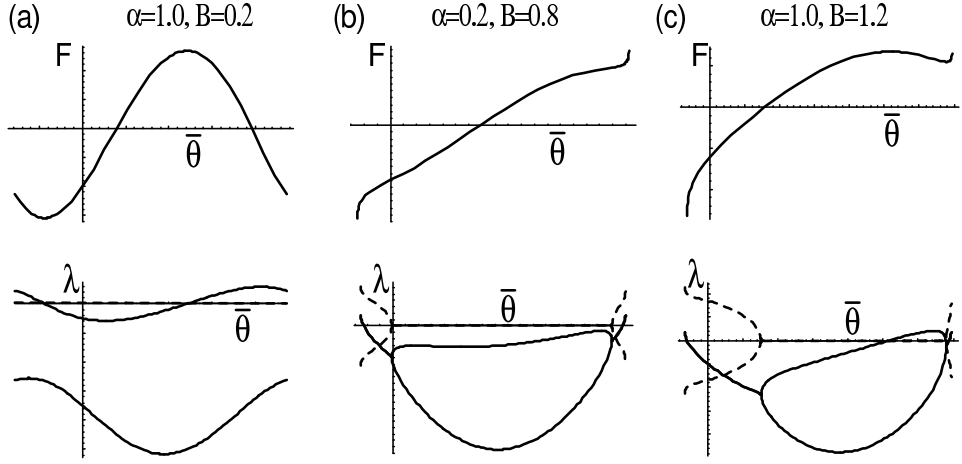}%
\caption{Behavior of the locking force $F(\bar{\theta})$ and the
stability eigenvalues $\lambda_{\pm}$ for (a) $\alpha=1.0,B=0.2$,
(b) $\alpha=0.2,B=0.8$, and (c) $\alpha=1.0,B=1.2$. The range of
angles plotted is $\bar{\theta}$ between $-\pi/2$ and $3\pi/2$
 for $B<2/\sqrt{27}$, panel (a), and between $-\bar{\theta}_{B}$
and $\pi+\bar{\theta}_{B}$ for $B>2/\sqrt{27}$, panels (b) and
(c). In the eigenvalue plots the solid curves are
$\operatorname{Re}\lambda_{\pm}$ and the dashed curves $\operatorname{Im}%
\lambda_{\pm}$.}%
\label{f_lambda}%
\end{center}
\end{figure*}
As $B$ increases, the behavior becomes quite complicated, and we have not
succeeded in proving any results about the full range of possible behavior of
$F(\bar{\theta})$. Some examples are shown in Fig.\ \ref{f_lambda}. For
$B<2/\sqrt{27}$ we have not uncovered parameters leading to an $F$-curve
qualitatively different from that in panel (a), i.e.\ a single maximum and
minimum. Note however, that for $B>1/\sqrt{8}$ values of $r<1/\sqrt{2}$ are
encountered, and so the Hopf bifurcation may limit the range of stable
solutions moving away from $\bar{\theta}=0$, before the maximum or minimum of
$F$ is reached. For $B>2/\sqrt{27}$ the physical solutions are limited to the
range $-\theta_{B}\leq\bar{\theta}\leq\pi+\theta_{B}$, and the slope of
$r(\bar{\theta})$ diverges at the end points: $dr/d\bar{\theta}\rightarrow
\infty$ for $\theta\rightarrow-\theta_{B}$ and $dr/d\bar{\theta}%
\rightarrow-\infty$ for $\theta\rightarrow\pi+\theta_{B}$. Since%
\begin{equation}
\frac{dF}{d\bar{\theta}}=-\frac{B}{r^{2}}\frac{dr}{d\bar{\theta}}(\alpha
\sin\bar{\theta}-\cos\bar{\theta})+\frac{B}{r}(\alpha\cos\bar{\theta}+\sin
\bar{\theta})
\end{equation}
we see that when $\left\vert dr/d\bar{\theta}\right\vert \rightarrow\infty$%
\begin{equation}
\frac{dF}{d\bar{\theta}}\rightarrow-\frac{F}{r}\frac{dr}{d\bar{\theta}}%
\end{equation}
so that $dF/d\bar{\theta}\rightarrow\pm\infty$ with the sign given
by the sign of $F$ for $\bar{\theta}\rightarrow\pi+\theta_{B}$
(where $dr/d\bar{\theta}<0$) and opposite to the sign of $F$ for
$\bar{\theta}\rightarrow-\theta_{B}$ (where $dr/d\bar{\theta}>0$).
This shows, for example, that for the parameters such as those in
panel (c) of Fig.\ \ref{f_lambda} where $F(\bar{\theta
}\rightarrow\pi+\theta_{B})>0$, the slope $dF/d\bar{\theta}$
approaches $+\infty$ as $\bar{\theta}\rightarrow\pi+\theta_{B}$.
This implies that either there is an additional minimum between
the maximum of $F$ and $\bar{\theta }\rightarrow\pi+\theta_{B}$ as
in this panel, or the maximum disappears and $F(\bar{\theta})$
becomes monotonically increasing in this region, as for the
parameters in panel (b). On the other hand for $\alpha=1,B=0.5$,
parameters between those of panels (a) and (c), it turns out that
$F(\pi+\theta_{B})<0$, so that
$dF/d\bar{\theta}\rightarrow-\infty$ as $\bar{\theta}\rightarrow
\pi+\theta_{B}$, and $F(\bar{\theta})$ may decrease monotonically
between the
maximum and $\bar{\theta}\rightarrow\pi+\theta_{B}$.%

\begin{figure}
[tbh]
\begin{center}
\includegraphics[width=2.8in]{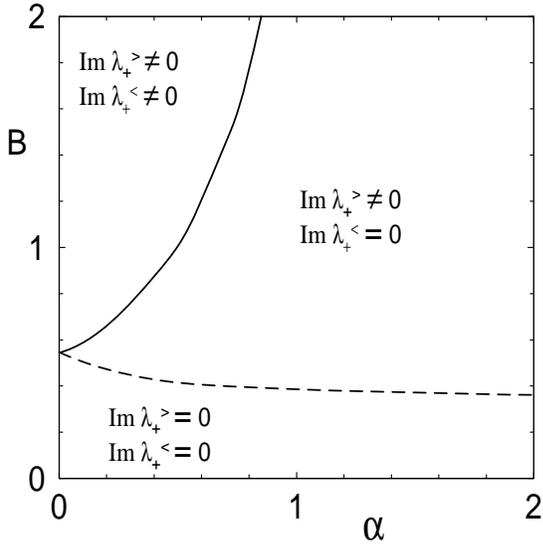}%
\caption{Regions of Hopf bifurcation ($\operatorname{Im}\lambda_{+}\neq0$) for
the first instability of the locked oscillator solution either side of
$\bar{\theta}=0.$ For $\bar{\theta}<0$ the value of $\operatorname{Im}%
\lambda_{+}$ is nonzero above the dashed curve, and for $\bar{\theta}>0$ it is
nonzero above the solid curve.}%
\label{Hopf_Stationary}%
\end{center}
\end{figure}
In Fig.\ \ref{Hopf_Stationary} the regions where the first
instability on either side of $\bar{\theta}=0$ is a Hopf
bifurcation ($\operatorname{Im}\lambda _{+}^{\lessgtr}\neq0$ when
$\operatorname{Re}\lambda_{+}^{\lessgtr}=0$, with
$\lambda_{+}^{<}$ the eigenvalue with larger real value for
$\bar{\theta}<0$, and $\lambda_{+}^{>}$ for $\bar{\theta}>0$) are
plotted as a function of $\alpha$ and $B$. Note that there may be
discontinuous jumps from $\operatorname{Im}\lambda_{+}=0$ to a
finite nonzero value of $\operatorname{Im}\lambda_{+}$: for
example on the negative $\bar{\theta}$ side, the minimum in $F$,
giving a stationary bifurcation, may disappear by colliding with
the maximum, and then the first instability jumps to the Hopf
bifurcation that was previously not the closest bifurcation to
$\bar{\theta }=0$. Figure \ref{Hopf_Stationary} does not tell the
whole story about whether the boundary of the locked state is a
stationary or Hopf bifurcation, because in general only the
instability on one side of $\bar{\theta}=0$ determines the
boundary, and which side this is depends on the distribution of
oscillator frequencies via the transverse self-consistency
condition. This is considered further below.

\subsection{Self consistency condition}%

\begin{figure}
[tbh]
\begin{center}
\includegraphics[width=2.8in]{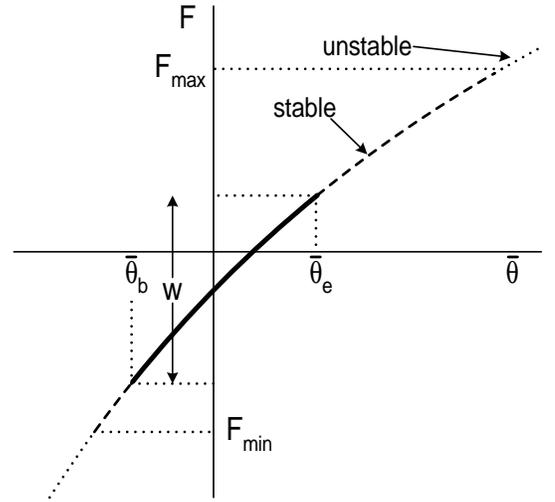}%
\caption{Schematic of transverse self consistency condition, showing the
locking force $F(\theta)$ acting on each oscillator as a function of the phase
relative to the order parameter $\theta$. The dashed portions of the curve
correspond to \emph{stable} solutions, and the dotted to \emph{unstable}. The
thick solid portion denotes the distribution of oscillator frequencies that
must be placed on the stable portion of $F$ and must also satisfy the
transverse consistency condition.}%
\label{Upper_lower}%
\end{center}
\end{figure}
We have found a range of $\bar{\theta}$ straddling $\bar{\theta}=0$ giving
stable locked solutions for individual oscillators. For a fully locked
solution, all the oscillators must have solutions to Eq.~(\ref{omega_F}) in
this range. In addition the imaginary part of the self consistency condition
Eq.~(\ref{ImSelfConsistency}) must be satisfied. This can be written in the
form%
\begin{equation}
\int\bar{g}(\bar{\omega})\,r(\bar{\omega})\sin[\bar{\theta}(\bar{\omega
})]\,d\bar{\omega}=0, \label{perp_int_omega}%
\end{equation}
since now there is a unique $\bar{\theta}$ and $r$ for each oscillator
frequency. Changing the integration variable in Eq.~(\ref{perp_int_omega}) to
the angle $\bar{\theta}$ yields%
\begin{equation}
\int\bar{g}(F(\bar{\theta}))\,r(\bar{\theta})\sin\bar{\theta}\left\vert
\frac{dF}{d\bar{\theta}}\right\vert \,d\bar{\theta}=0. \label{perp_int}%
\end{equation}
The degree of freedom, for fixed $\alpha,B$ and width of frequency
distribution $w$, to be determined by this condition is the order parameter
frequency $\Omega$. The scheme of imposing the condition is sketched in
Fig.~\ref{Upper_lower}. The thick solid line, of fixed length $w$ along the
$F$ axis, is to be slid along the curve $F(\bar{\theta})$ (corresponding to
varying $\Omega$) until the integral Eq.~(\ref{perp_int}) is zero. All
quantities, except $\sin\bar{\theta}$ are positive, and so this line must
straddle the origin. This must be done with all of the solid line lying within
the stable (dashed) range of $F(\bar{\theta})$.

For a bounded frequency distribution we can always find a fully locked
solution for large enough $\alpha\beta$. The argument is as follows. First,
there is always a range of stable locked oscillator solutions straddling
$\bar{\theta}=0$. If we define $\omega=0$ as the center of the distribution
$g(\omega)$ then the order parameter frequency is given by%
\begin{equation}
\Omega=-F_{c}-\alpha-B
\end{equation}
where $F_{c}$ is $F(\bar{\theta})$ evaluated at the center (with
respect to the ordinate) of the solid portion of the curve in
Fig.~\ref{Upper_lower}, once Eq.~(\ref{perp_int}) is satisfied.
Since it follows from Eqs.~(\ref{F_bar_theta}) and
(\ref{rConstant}) that the slope of $F(\bar{\theta})$
at $\bar{\theta}=0$ is%
\begin{equation}
F^{\prime}(0)=B(\alpha+B/2),
\end{equation}
the range of phase angles of the locked oscillators of order $w/F^{\prime}(0)$
is small for large enough $B$ or $\alpha B$, and there is always a fully
locked solution for a bounded frequency distribution in this limit. The center
of the band can then be evaluated as $\omega=0$, and so%
\begin{equation}
\Omega=-F(0)-\alpha-B.
\end{equation}
Also, in this limit all the oscillators have essentially the same phase, so
that $R=1$ and $\beta=B$. This gives the complete solution for the fully
locked state for very large $\beta$.

It is easiest to understand the limit of the fully locked solution by
decreasing $B$ at fixed $\alpha,w$. As $B$ decreases, the range of the locking
force $F$ decreases. We can continue to construct the solution as in
Fig.~\ref{Upper_lower}, with the portion of the $F(\bar{\theta})$ covered by
the locked oscillators determined by the transverse self consistency
condition, until the first value $B=B_{c}$ is reached for which \emph{either}
the lower end of the locked band would pass below $F_{\min}^{(s)}$, \emph{or}
the upper end of the locked band would pass beyond $F_{\max}^{(s)}$. This
signals the onset of instability of the corresponding locked oscillator,
either by a stationary or Hopf bifurcation depending on $\operatorname{Im}%
\lambda_{+}$ at the appropriate $F_{\min}$ or $F_{\max}$
\footnote{For some $\alpha$ $F_{\min}$ or $F_{\max}$ may jump
discontinuously as a function of $B$ if a new bifurcation occurs
when $dF/d\bar{\theta}\rightarrow0$, and a new minimum and
maximum of $F(\bar{\theta})$ are created at a previously interior
point of the stable band. In such a case there may be no value of
$B$ for which one end of the band coincides with $F_{\min}$ or
$F_{\max}$ signaling the onset of instability. Presumably this
corresponds to a first order transition out of the fully locked state.}.%

\begin{figure}
[tbh]
\begin{center}
\includegraphics[width=3.3in]{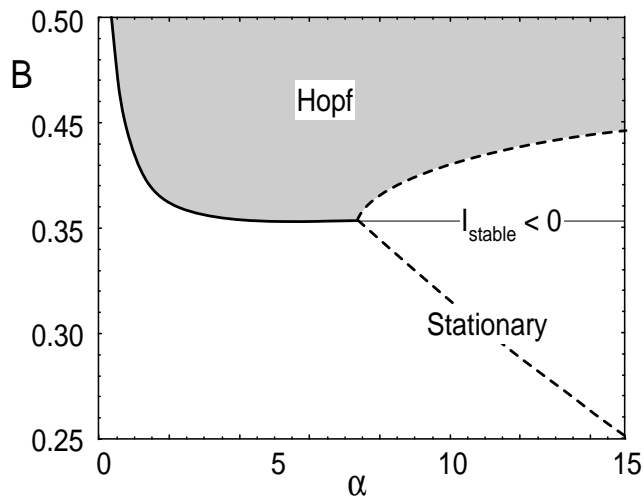}%
\caption{Plot showing regions in the $B-\alpha$ plane for which
the instability from the fully locked state is Hopf (shaded) or
stationary (unshaded) for a top-hat distribution. The dashed line
shows the range of $\alpha,B$ for which $I_{\mathrm{stable}}<0$
for a top-hat distribution. (Note the change of range of
$\alpha,B$ plotted compared with
Fig.~\ref{Hopf_Stationary}.)}%
\label{hopf_stationary_1}%
\end{center}
\end{figure}
We could also imagine increasing the width of the distribution $w$ from the
fully locked solution which occurs at small $w$ for fixed $\alpha$ and $B$,
until one end of the growing band of locked oscillator solutions reaches
$F_{\min}$ or $F_{\max}$. For a top-hat distribution of oscillator frequencies
the integral%
\begin{equation}
I_{\mathrm{stable}}=\int_{\mathrm{stable}}\bar{g}(F(\bar{\theta}%
))\,r(\bar{\theta})\sin\bar{\theta}\left\vert \frac{dF}{d\bar{\theta}%
}\right\vert \,d\bar{\theta},
\end{equation}
where the integration extends over the whole stable range of $\bar{\theta}$,
provides an indicator of whether the limit is reached at the lower or upper
bound of the stable band: if $I_{\mathrm{stable}}$ is \emph{positive}, then to
satisfy Eq.~(\ref{perp_int}) the integral must be \emph{reduced} by
\emph{lowering} the \emph{upper} integration bound, and so the range of
integration extends from the \emph{lower} stability bound. This gives the
condition for the maximum $w$%
\begin{equation}
w_{c}/2=F_{c}-F_{\min}^{(s)},
\end{equation}
with $F_{\min}^{(s)}$ the lowest $F$ for stable solutions and
$F_{c}$ the value of $F$ at the center of the band for
$I_{\mathrm{stable}}=0$. On the other hand if
$I_{\mathrm{stable}}$ is negative, then the integral
Eq.~(\ref{perp_int}) must be \emph{increased} by \emph{raising}
the \emph{lower} integration bound, and so the range of
integration extends to the \emph{upper} stability bound.
This gives the condition%
\begin{equation}
w_{c}/2=F_{\max}^{(s)}-F_{c},
\end{equation}
with $F_{\max}^{(s)}$ the largest $F$ for stable solutions. We can then find
the range of $\alpha,B$ for which the locked state disappears by a Hopf or by
a stationary bifurcation, Fig.~\ref{hopf_stationary_1}. This is constructed
from Fig.~\ref{Hopf_Stationary}, giving the conditions for the instabilities
at $F_{\min}^{(s)}$ and $F_{\max}^{(s)}$ to be Hopf or stationary, and the
result just determined for which of these instabilities limits the range of
the fully locked solution. Numerical results show that for the top-hat
distribution the condition $I_{\mathrm{stable}}<0$ occurs only for a
restricted range of parameters: large $\alpha$ and $B$ near $2/\sqrt{27}$, the
region to the right of the dashed line in Fig.\ \ref{hopf_stationary_1}.
Combining these results yields a Hopf bifurcation from the fully locked state
in the shaded region of Fig.~\ref{hopf_stationary_1}. For other oscillator
distribution shapes we do not have a criterion for the nature of the
instability without a detailed solution of the self consistency condition for
each width.

So far the solution has been developed in terms of $\alpha,B$. We
now determine the magnitude $R$ of the order parameter from the
parallel self-consistency condition, the real part of
Eq.~(\ref{SelfConsistency})
which can be written in the form%
\begin{equation}
\int_{\bar{\theta}_{b}}^{\bar{\theta}_{e}}\bar{g}(F(\bar{\theta}%
))\,r(\bar{\theta})\cos\bar{\theta}\left\vert \frac{dF}{d\bar{\theta}%
}\right\vert \,d\bar{\theta}=R,
\end{equation}
where the range of integration is that determined from the
transverse self-consistency condition, see Fig.~\ref{Upper_lower}.
From $R(\alpha,B)$ we can calculate $\beta=B/R$ at the boundary of
locked solutions. If the dependence $R(B)$ is smooth and
monotonic, we can map the results depending on $(\alpha,B)$ onto
functions of $(\alpha,\beta).$ However, discontinuities in $R(B)$
might well occur due to the jumps in the stability range, for
example when the stationary bifurcation disappears as described
before. This might lead to values of $\beta$ for which no
prediction, e.g.\ for $w_{c}(\alpha,\beta)$, has been yielded by
the
algorithm. We do not have results for such cases.%

\begin{figure}
[tbh]
\begin{center}
\includegraphics[width=2.8in]{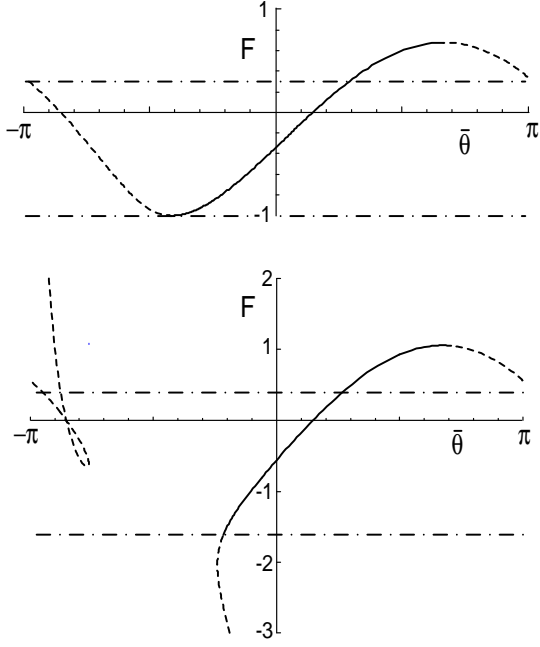}%
\caption{Plots of $F(\bar{\theta})$ for $\alpha=2$, $B=0.34$, $w=1.3$ (top
panel), and $\alpha=2$, $B=0.56$, $w=2$ (lower panel) corresponding to the
boundary of the fully locked solution: solid curve - stable solutions; dashed
curve - unstable solutions. The dash-dotted lines straddle the range of
$\bar{\omega}=F(\bar{\theta})$ of locked oscillators, between $F_{\min}$ and
$F_{\min}+w$. The values of the order parameter are 0.92 (upper panel) and
0.93 (lower panel).}%
\label{TwoBeta}%
\end{center}
\end{figure}
Two examples of the construction of locked solutions for the
triangular distribution, defined below in Eqs.~(\ref{triangular}),
are shown in Fig.~\ref{TwoBeta}: the first for small $B=\beta R$,
where a solution $r(\bar{\theta})$ exists for all $\bar{\theta}$;
and the second for larger $B$ where there are ranges of
$\bar{\theta}$ for which no physical solution for the magnitude
(i.e.~real and positive) exists. The value of $B$ for which Eq.
(\ref{Transverse}) is satisfied in each case was found by simple
bisection applied to the numerical results of the integration. In
both the cases shown the boundary of the fully locked region
occurs when $\bar{\omega}_{\min }=F_{\min}^{(s)}$. For the smaller
value of $B$ this condition corresponds to the limit of the
existence range of solutions. For the larger value of $B$ the
limit occurs at the instability of the locked oscillator solution,
which develops via a Hopf bifurcation
($\operatorname{Im}\lambda_{+}\neq0$). Results for the boundary of
the fully locked state for a top-hat distribution with $g(0)=1$
are shown in Fig.~\ref{th_pd}, and for the triangular distribution
with $g(0)=1$ in Fig.~\ref{tri_pd}.

\section{Results for various frequency distributions}

\label{sec_results}

\subsection{Lorentzian Distribution}
\label{sec_lort}
In this section we present detailed results for the case of a Lorentzian
distribution of frequencies. We concentrate on a distribution with $g(0)=1$,
but also present some results for a wider distribution $g(0)=\frac{1}{2}$,
which shows some novel features. We begin this section by calculating the
linear stability of the unsynchronized state. Since the Lorentzian
distribution is unbounded, there is no fully locked solution.

For the purposes of analysis we choose, without loss of generality, a
Lorentzian distribution $g(\omega)$ centered about zero frequency%
\begin{equation}
g(\omega)=g(0)\frac{w^{2}}{\omega^{2}+w^{2}}\qquad\text{with}\qquad
w=(\pi g(0))^{-1}.
\label{lorentzian_analytic}%
\end{equation}
The half width at half height is $w$. In terms of the shifted frequency
$\bar{\omega}$ the distribution is%
\begin{equation}
\bar{g}(\bar{\omega})=g(0)\frac{w^{2}}{(\bar{\omega}+\delta)^{2}+w^{2}},
\end{equation}
with%
\begin{equation}
\delta=\alpha+\beta+\Omega. \label{define_delta}%
\end{equation}

The integrals Eqs.~(\ref{integrals}) are%
\begin{subequations}
\begin{align}
I_{1}  &  =\frac{2+w}{2(4+4w+w^{2}+\delta^{2})},\\
I_{2}  &  =\frac{\delta}{w^{2}+\delta^{2}},\\
I_{3}  &  =\frac{\delta}{4+4w+w^{2}+\delta^{2}},\\
I_{4}  &  =\frac{w}{w^{2}+\delta^{2}}.
\end{align}
The imaginary part of the self consistency condition Eq.~(\ref{perp_integrals}%
) reduces to%
\end{subequations}
\begin{equation}
\delta^{2}-2\alpha\delta+2w+w^{2}=0.
\end{equation}
This serves to fix the frequency of the order parameter $\Omega$ at the onset
of synchronization via Eq.~(\ref{define_delta}) in terms of the parameters of
the system $\alpha,\beta,w$. There are \emph{two} solutions for $\delta$%
\begin{equation}
\delta=\alpha\pm\sqrt{\alpha^{2}-(2w+w^{2})}. \label{delta}%
\end{equation}
For large $\alpha$ or small $w$ the approximate solutions are
$\delta \simeq(w+\frac{1}{2}w^{2})/\alpha$ giving a locking
frequency near the center of the band of the free running
oscillators, and $\delta\simeq2\alpha$ giving a locking frequency
far in the tails. Note that the requirement that $\delta$ is real
means that $\alpha$ must be sufficiently large $\left\vert
\alpha\right\vert >\alpha_{\min}$ with%
\begin{equation}
\alpha_{\min}=\sqrt{w^{2}+2w}. \label{eq_amin}%
\end{equation}
For $0<\left\vert \alpha\right\vert <\alpha_{\min}$ the unsynchronized state
is linearly stable for all values of $\beta$.

The critical value of $\beta$ is determined from Eq.~(\ref{par_integrals}) and
evaluates to%
\begin{equation}
\beta_{c}=\frac{\left(  w^{2}+{\delta}^{2}\right)  \,\left(  4+4\,w+w^{2}%
+{\delta}^{2}\right)  }{(2\,w\,+w^{2})\left(  \alpha+\delta\right)
+\delta\,\left(  2-\alpha\,\delta+{\delta}^{2}\right)  }, \label{betac}%
\end{equation}
where the expression Eq.~(\ref{delta}) for $\delta$ is to be
substituted. Given a width $w$ of the oscillator distribution, for
each $\alpha
>\alpha_{\min}(w)$ there are two critical values of $\beta$: $\beta_{c-}$ and
$\beta_{c+}$ [corresponding to the minus and plus signs in the
expression Eq.~(\ref{delta}) for $\delta$], such that the
unsynchronized state is \emph{unstable} for
$\beta_{-}<\beta<\beta_{+}$. It is remarkable that for very strong
coupling, $\beta$ large, the unsynchronized state remains stable.
However, as we have already seen and will discuss in more detail,
a large amplitude synchronized state is also stable in this
regime. For $\alpha\gg\alpha_{\min}$ the results for $\beta_{c}$
reduces to $\beta_{c}\simeq w(w+2)\alpha^{-1}$ and $4\alpha$. In
the limit $w\ll1$ the former result reproduces the result
$\alpha\beta_{c}\rightarrow2/\pi g(0)$ expected from the reduction
to the phase equation valid in this limit.

We have used simulations to confirm the boundaries for instability of the
unsynchronized state as well as to study the behavior subsequent to the
instability. The numerical results use the cutoff Lorentzian form introduced
in Eq.~(\ref{eq_para_lort}). Two widths are considered: a narrow one with a
peak height of $g(0)=1$ and one approximately twice as wide with $g(0)=1/2$.
In all cases the distribution tails are removed above some large frequency.
Qualitative differences in the phase diagrams are observed for the two
distribution widths.
%

\begin{figure}
[tbh]
\begin{center}
\includegraphics[width=3.3in]{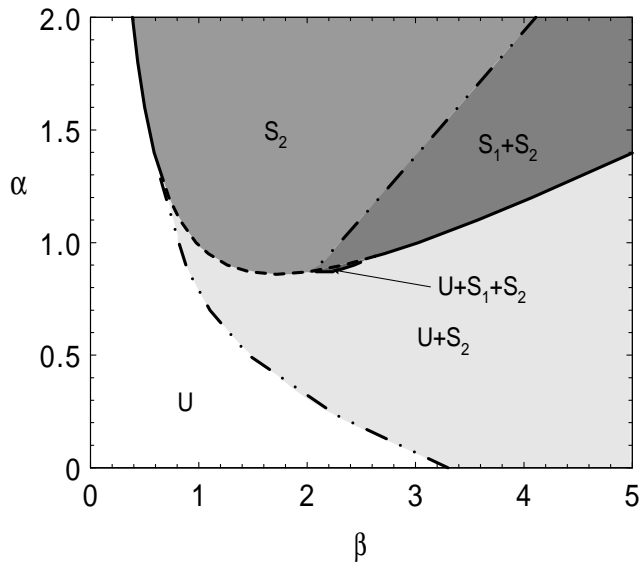}%
\caption{Phase diagram for a Lorentzian distribution of
frequencies with width such that $g(0)=1$. Solid and dashed lines
show analytical results of the linear stability of the
unsynchronized state. Numerics show the bifurcations are
supercritical along the solid portions and subcritical along the
dashed portion. Dash-dotted lines are saddle-node bifurcations
observed in numerical simulations. States are: $U$ -
unsynchronized; $S_{1},S_{2}$ synchronized with small and large
amplitude
respectively.}%
\label{LorentzianPD}%
\end{center}
\end{figure}

The phase diagram for the (narrow) Lorentzian distribution with
$g(0)=1$ is shown in Fig.~\ref{LorentzianPD}. The solid and dashed
lines are the analytically obtained stability boundaries of the
unsynchronized solution. The numerical simulations show that over
the dashed portion of the linear instability curve the bifurcation
is subcritical, giving a jump in the order parameter magnitude $R$
at onset. In addition the sweeps yield a number of saddle-node
bifurcations identified as discontinuous jumps in $R$; these are
denoted by dash-dotted lines in Fig.~\ref{LorentzianPD}; we do not
have closed form relations for these boundaries. Thus along the
dashed or dash-dotted portions of the boundaries, discontinuous
jumps in $R$ occur, either between the unsynchronized state and a
synchronized state, or between two synchronized states with
different values of $R$. The two synchronized states are labelled
$S_{1}$ and $S_{2}$ in Fig.~\ref{LorentzianPD}. When they coexist
at the same $\alpha$ and $\beta$ the state $S_{2}$ has the larger
value of $R$, but for both states $R$ may go to zero, connecting
continuously
with the unsynchronized state $U$, for some values of $\alpha,\beta$.%

\begin{figure}
[tbh]
\begin{center}
\includegraphics[width=3.3in]{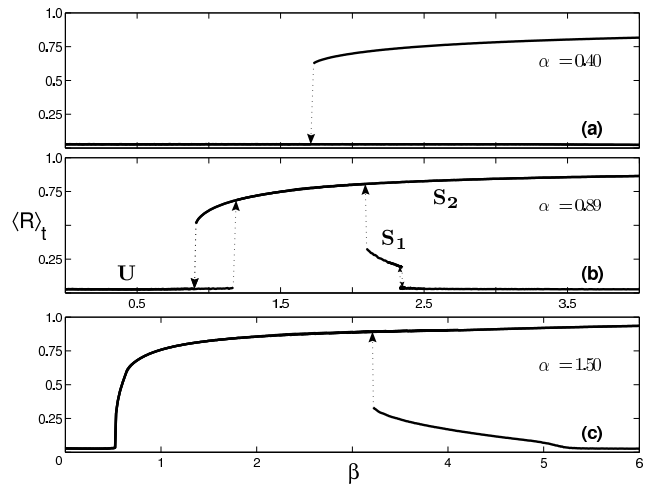}%
\caption{Slices of the phase diagram Fig.~\ref{LorentzianPD}
showing the time averaged order parameter magnitude $\langle
R\rangle_{t}$ as a function of $\beta$ from numerical simulations
of the cutoff Lorentzian distribution with $g(0)=1$. In panel (a)
the only solutions are the unsynchronized ($U$) and large $R$
synchronized ($S_{2}$) states, while in panels (b) and (c) the
small $R$ synchronized solution is also stable. Arrows denote
discontinuous jumps that were observed as
simulations followed the various solution branches.}%
\label{fig_cuts_ahalf_d8}%
\end{center}
\end{figure}
Representative phase diagram slices for the Lorentzian case with
$g(0)=1$, showing the time-averaged order parameter magnitudes
$\langle R\rangle_{t}$ as a function of $\beta$ at fixed $\alpha$,
are presented in Fig.~\ref{fig_cuts_ahalf_d8}. In agreement with
Eq.~(\ref{eq_amin}) the unsynchronized state is stable for all
$\beta$ provided $\alpha<0.872$. Simulations find the $S_{2}$
solution is bistable with $U$ over this $\alpha$ range at larger
$\beta$ values, as shown in Fig.~\ref{fig_cuts_ahalf_d8}(a). For
$\alpha>\alpha_{min}$ the unsynchronized state is unstable over a
range in $\beta$. As shown in Fig.~\ref{fig_cuts_ahalf_d8}(b) for
a range of $\alpha$ near $\alpha_{\min}$ a subcritical bifurcation
occurs at $\beta_{c-}$ as the $U$ solution becomes unstable and
$S_{2}$ forms. With increasing $\beta$ $S_{2}$ is the only stable
solution until a region of bistability where both synchronized
solutions coexist. As shown in the phase diagram there is a small
region (labelled $U+S_{1}+S_{2}$) over which all three solutions
are simultaneously stable. This region of tristability can be
observed in Fig.~\ref{fig_cuts_ahalf_d8}(b), over the $\beta$
range of hysteric transition between $S_{1}$ and $U$ near
$\beta=2.3$. With increasing $\alpha$ the system passes through a
tricritical point where the subcritical bifurcation at
$\beta_{c-}$ becomes supercritical. The tricritical point has a
codimension of 2. An example slice at $\alpha$ sufficiently large
that the bifurcation is supercritical is shown in Fig.~\ref{fig_cuts_ahalf_d8}%
(c), where with increasing $\beta$ regions of bistability between the
synchronized solutions and then $U$ and $S_{2}$ are observed.

\begin{figure}
[tbh]
\begin{center}
\includegraphics[width=3.3in]{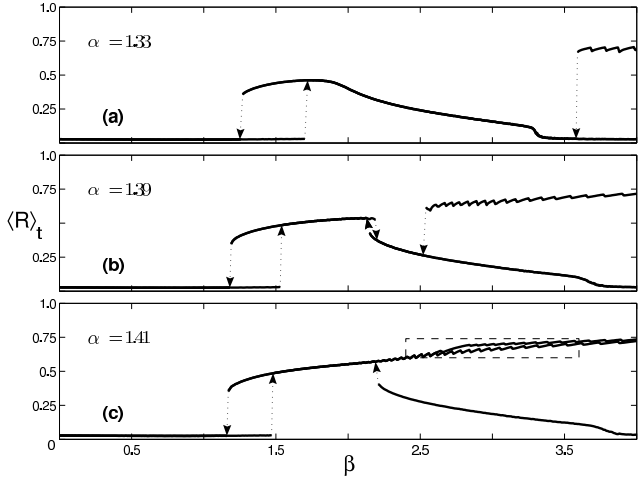}%
\caption{Constant $\alpha$ slices from numerical simulations of
the cutoff wide Lorentzian with $g(0)=1/2$. Shown is the time
averaged order parameter magnitude $\langle R\rangle_{t}$ for a
range of $\beta$ at three $\alpha$ values: (a) $\alpha=1.33$, (b)
$\alpha=1.39$, and (c) $\alpha=1.41$. These solutions are observed
in an array of $N=1000$ oscillators with a Lorentzian frequency
distribution with $g(0)=1/2$ and a cutoff frequency of
$\omega_{c}=16$. The dashed
box region in panel (c) is examined more closely in Fig.\ \ref{fig_zig_zag}.}%
\label{fig_soln_cuts_a1_d16}%
\end{center}
\end{figure}

We now consider a Lorentzian frequency distribution of
approximately twice the width. Specifically, we take $g(0)=1/2$ in
Eq.~(\ref{eq_para_lort}) or Eq.~(\ref{lorentzian_analytic}). There
are a number of changes to the details of the phase diagram that
we do not discuss in detail. Several of these are straightforward
consequences of the wider frequency spread: for example the
instability of the unsynchronized solution moves to larger
$\alpha$ and $\beta$. Instead we focus on two particular features.

The first new feature that is evident from the fixed $\alpha$ cuts
shown in Fig.\ \ref{fig_soln_cuts_a1_d16} is the reconnection of
the branches of synchronized solutions that occurs as $\alpha$ is
decreased. In Fig.\ \ref{fig_soln_cuts_a1_d16}(c), typical of
larger values of $\alpha$, the synchronized state growing from the
$\beta_{c+}$ instability ends at a saddle-node bifurcation, with
the order parameter jumping to larger values as $\beta$ is
decreased. This is the same as the behavior for the narrower
distribution, Fig.\ \ref{fig_cuts_ahalf_d8}. On the other hand for
smaller values of $\alpha$, as in Fig.\
\ref{fig_soln_cuts_a1_d16}(a), this state merges continuously with
the larger magnitude state. This change in the topology of the
solution branches as $\alpha$ increases occurs through the
development of two additional saddle-nodes, as shown in Fig.\
\ref{fig_soln_cuts_a1_d16}(b) near $\beta=2.1$, and the collision
of one of these with the saddle-node terminating the large-$\beta$
upper branch.%

\begin{figure}
[tbh]
\begin{center}
\includegraphics[width=3.3in]{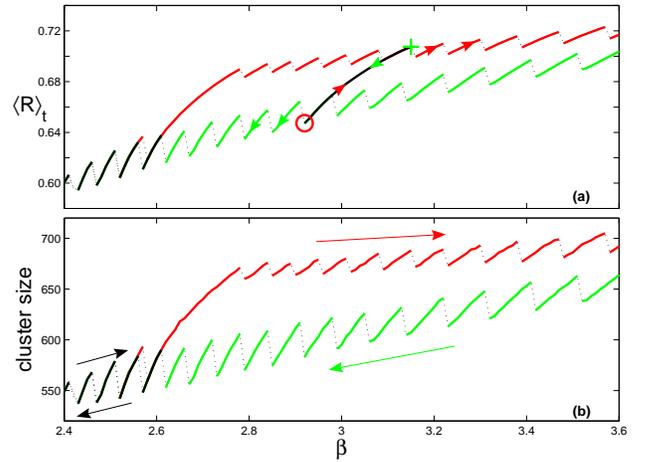}%
\caption{Hysteretic behavior between two boundaries of the large
magnitude $\langle R\rangle_{t}$ synchronized solution for a
Lorentzian frequency distribution (color online).  Shown are the
results for $N=1000$, $\alpha=1.41$, and a cutoff Lorentzian
distribution with $g(0)=1/2$ and $\omega_{c}=16$. The arrows
denote the direction of the $\beta$ sweep, and the states are
colored red for an upward sweep, green for a downward sweep, and
black if both directions lead to the same state. The + symbols
denote the starting point for particular sweeps to traverse a
state interior to the band (green for the start of a downward
sweep, red for an upward sweep).  Panel (a) is a magnification of
the dashed box in Fig.~\ref{fig_soln_cuts_a1_d16}(c) showing the
$\langle R\rangle_{t}$ over a range of $\beta$.  In (b) the
corresponding size of the synchronized cluster measured by the
synchronization index is
shown.}%
\label{fig_zig_zag}%
\end{center}
\end{figure}

Another apparent difference with the wider frequency distribution
is the steps displayed by the large amplitude synchronized state
at larger $\beta$ in Fig.~\ref{fig_soln_cuts_a1_d16}. While these
steps also occur with the narrow frequency distribution, their
increased amplitude with the wider frequency distribution
facilitates the presentation of the behavior. As an example, the
dashed region in Fig.~\ref{fig_soln_cuts_a1_d16}(c) is shown
enlarged in Fig.~\ref{fig_zig_zag}. Panel (a) in this figure plots
the time averaged order parameter magnitude $\langle
R\rangle_{t}$, while (b) shows that the solution changes correlate
exactly with changes in the number of oscillators locked to the
synchronized cluster.

Interestingly, the solution follows different trajectories
depending on the history of changing $\beta$. For example, if
$\beta$ is monotonically increased the system settles onto the
upper trajectory of solution steps (colored red) while the lower
trajectory of solution steps (colored green) is observed when
$\beta$ is monotonically decreased. These solutions overlap at
smaller $\beta$ (colored black).

The two solution trajectories are in fact the boundaries of a band
of multistable solution branches. This structure is demonstrated
in Fig.~\ref{fig_zig_zag}(a) by beginning the system on each of
these two boundaries and reversing the direction of $\beta$
variation. Now as $\beta$ changes the solution moves off its
boundary along a continuous line to the other boundary where it
follows the expected path for that sense of $\beta$ changes.
Starting solutions are represented by the + symbols colored for
the direction that $\beta$ is to be varied. Beginning on the lower
boundary, that is observed when decreasing $\beta$, and increasing
$\beta$ from the lower (red) + the oscillators follow the solid
(black) line in the direction of the upward arrow to the upper
boundary. Likewise, beginning on the upper boundary, that is
observed when increasing $\beta$, and decreasing $\beta$ the
oscillators display a series of states beginning with the upper
(green) + and follow the solid (black) line to the lower boundary
in the direction of the downward arrow. In this way, the
synchronized solution with large time-averaged order parameter
magnitude turns out to be a band of solutions, with striations
that can be accessed by appropriate changes in $\beta$. The same
is true for the narrow Lorentzian, however the band is smaller.

The steps in the magnitude of the order parameter are found to be associated
with jumps in the number of oscillators locked to the phase of the order
parameter. Defining a synchronized cluster as the oscillators locked to
$\Theta$ we find the discrete steps in $R$ are due to a small group of
oscillators leaving the cluster, and then recombining with the cluster one at
a time as $\beta$ increases. In Fig.~\ref{fig_zig_zag}(b) the cluster size is
measured by the phase synchronization index, defined by Eq.~\ref{eq_sync_ind},
as the number of oscillators with $\chi_{n}=1-\epsilon$, where $\epsilon$ is
some small number, $\epsilon<<1$, to allow for some phase variation over the
finite time of measurement. With increasing system size any small group of
oscillators leaving the cluster can be expected to have decreasing influence
on the order parameter.

As the number of oscillators tends to infinity the solution band
becomes a region densely populated by these striations as the
discrete steps become closer and decrease in length. For the
simulations in Fig.~\ref{fig_zig_zag} $N=1000$ oscillators were
used. To investigate finite size effects we also studied this
solution using $N=5000$ and $N=10,000$. With increasing system
size the individual steps on each solution boundary occur over a
more narrow range in $\beta$, becoming shorter in length, and
correspondingly more densely packed. Thereby, while the discrete
steps will disappear in the infinite size limit the width of the
band of synchronized solutions remains nearly the same. In this
limit synchronized states will move along one of the striations to
the band boundary appropriate for the direction of the $\beta$
sweep.

\subsection{Top-hat Distribution}

\label{sec_top} For bounded distributions we can calculate the linear
stability boundaries of both the unsynchronized and the fully synchronized
states. We first do this for a uniform bounded distribution, which we call a
top-hat distribution.

A top hat distribution centered on $\omega=0$ is given by%
\begin{equation}
g(\omega)=\left\{
\begin{tabular}
[c]{cc}%
$w^{-1}$ & for $\left\vert \omega\right\vert \leq w/2$\\
$0$ & for $\left\vert \omega\right\vert >w/2$%
\end{tabular}
\ \ \ \ \right.
\end{equation}
The integrals Eqs.~(\ref{integrals}) for this distribution are%
\begin{subequations}
\begin{align}
I_{1}  &  =\frac{1}{2w}\tan^{-1}\left[  \frac{w/2}{(1+\delta^{2}/4)-w^{2}%
/16}\right]  ,\\
I_{2}  &  =\frac{1}{w}\ln\left\vert \frac{w-2\delta}{w+2\delta}\right\vert ,\\
I_{3}  &  =\frac{1}{2w}\ln\left[  \frac{1+(w-2\delta)^{2}/16}{1+(w+2\delta
)^{2}/16}\right]  ,\\
I_{4}  &  =\left\{
\begin{tabular}
[c]{cc}%
$\frac{\pi}{w}$ & for$~\delta<w/2$\\
$0$ & for $\delta>w/2$%
\end{tabular}
\ \ \ \right.  .
\end{align}
The self consistency equations (\ref{perp_integrals}) and (\ref{par_integrals}%
) can be solved numerically. The equations simplify in the limit of small $w$,
and the results here can be displayed in closed form.

In small $w$ limit there is one solution of Eq.~(\ref{perp_integrals}) giving
a locked frequency within the band of (shifted) oscillator frequencies,
$\delta<w/2$, for which $I_{1}$ and $I_{3}$ may be neglected in the small $w$
limit. Equation (\ref{perp_integrals}) then gives the explicit expression for
the frequency offset from mid band%
\end{subequations}
\begin{equation}
\delta\simeq\frac{w}{2}\tanh\left(  \frac{\pi}{2\alpha}\right)  ,
\label{u-approx1}%
\end{equation}
and Eq.~(\ref{par_integrals}) gives the condition of the parameters at the
onset of synchronization%
\begin{equation}
(\alpha+\alpha^{-1})\beta_{c}\simeq\frac{2w}{\pi}. \label{u-approx2}%
\end{equation}
Notice that for large $\alpha$ the locking frequency is close to the center of
the band, and the critical condition reduces to%
\begin{equation}
\alpha\beta_{c}=2w/\pi,
\end{equation}
the value expected from the mapping onto the Kuramoto model for
large $\alpha $. For small $\alpha$ on the other hand $\delta$
approaches $w/2$ giving us the result that the locking frequency
approaches the upper edge of the band. In this case the onset
occurs at $\beta_{c}\simeq2w\alpha/\pi$. Even for moderate values
of $\alpha$ such as $\alpha=1$, the locking frequency is far off
the band center (a fraction $0.92$ of the half band width for
$\alpha=1$).

The second solution to Eq.~(\ref{perp_integrals}) in the small $w$ limit gives
an order parameter frequency \emph{outside} of the band, $\delta>w/2$. For
small $w$ the frequency is given by%
\begin{equation}
\delta\simeq\frac{w}{2}\coth\left(  \frac{w}{4\alpha}\right)  ,
\label{u-approx3}%
\end{equation}
and the onset of synchronization occurs at%
\begin{equation}
\beta_{c}\simeq4\alpha. \label{u-approx4}%
\end{equation}
The solution that grows from this instability is a remarkable
state in that it is \emph{synchronized} in the sense that there is
a nonzero value of the order parameter $\left\vert \psi\right\vert
\neq0$, but there are no oscillators \emph{frequency locked} to
one another or to the frequency of the order parameter: a plot of
the actual frequency distribution as in
Figs.~\ref{Fig_smr_plateau}(a) and \ref{Fig_lgr_plateau}(a) shows
a smooth curve with no plateau. Numerical investigation of this
state shows that instead, the distribution of oscillators
$\rho(\omega,\theta)$ is enhanced for phases near the phase of the
order parameter: oscillators slow down in this vicinity (i.e.\
$d\bar{\theta}/dt$ becomes smaller), but do not come to rest
relative to the order parameter. Now the ordered oscillator
frequency plot is
continuous, with no plateau of locked oscillators.%
\begin{figure}
[tbh]
\begin{center}
\includegraphics[width=3.3in]{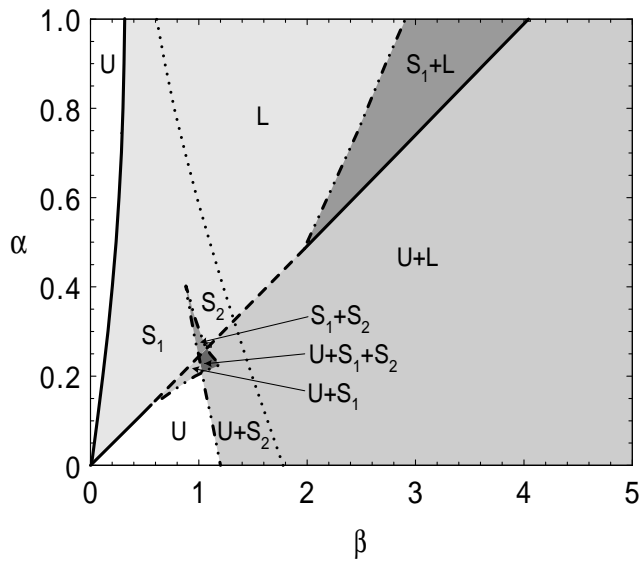}%
\caption{Phase diagram for a top-hat distribution of frequencies
of width $w=1$ such that $g(0)=1$. The solid and dashed lines are
from the linear stability analysis of the unsynchronized state:
the unsynchronized state is unstable for $\beta$ values between
the two portions of these lines. The numerical results show that
the solid portion is supercritical, whereas the dashed portion is
subcritical. Dash-dotted lines are saddle-node bifurcations from
numerical simulations. The dotted line is the stability boundary
of the fully locked solution calculated using the methods of Sec.
\ref{Sec_FullLocking}. States are as in Fig.\ \ref{LorentzianPD}
with, in addition, $L$ denoting the fully locked state.}%
\label{th_pd}%
\end{center}
\end{figure}
The linear instability boundaries for a top-hat distribution of full width
$w=1$ (giving $g(0)=1$) are shown as the solid and dashed lines in
Fig.\ \ref{th_pd}. The approximate solutions Eqs.~(\ref{u-approx1}%
-\ref{u-approx4}) turn out to be quite accurate even for $w=1$:
the approximate curves are indistinguishable from the numerical
curves in Fig.\ \ref{th_pd}. Note that unlike the case of the
Lorentzian distribution, the linear instability persists for
$\alpha,\beta\rightarrow0$. In this limit the frequency of the
order parameter is right at the edge of the distribution of
oscillator frequencies. The discontinuity of the distribution of
oscillator frequencies at the band edge seems to be responsible
for the persistence of
the instability to small values of the coupling constants.%

\begin{figure}
[tbh]
\begin{center}
\includegraphics[width=3.3in]{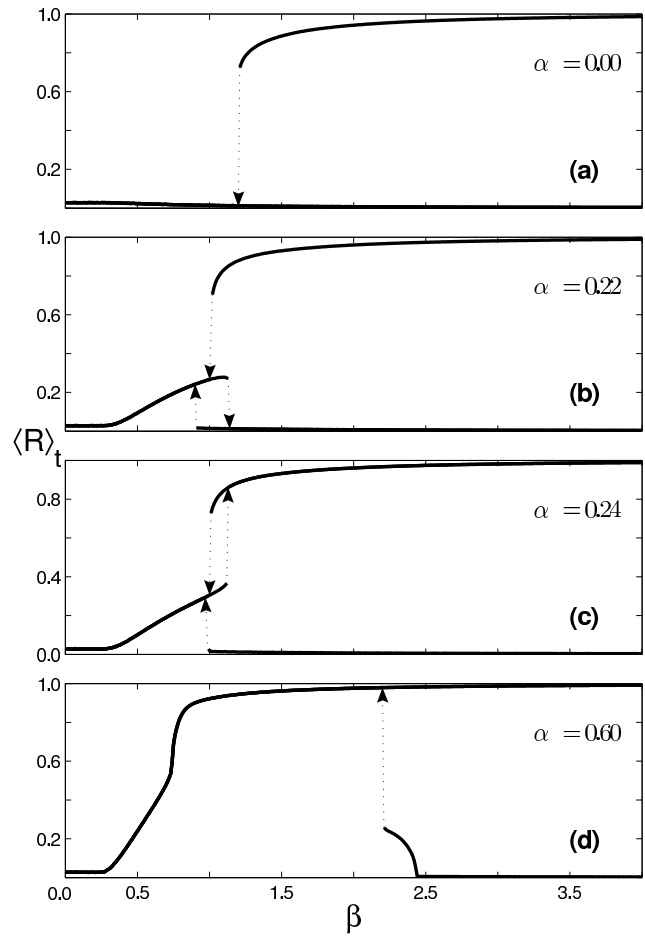}%
\caption{Phase diagram slices observed in simulations with a
top-hat frequency distribution. The time averaged order parameter
magnitude $\langle R\rangle_{t}$ over a range of $\beta$ is shown
for constant $\alpha$: (a) $\alpha=0$, (b) $\alpha=0.22$, (c)
$\alpha=0.24$, and (d) $\alpha=0.60$. In
these simulations $N=1000$ and the width $w=1$.}%
\label{fig_cuts_tophat}%
\end{center}
\end{figure}
The boundary of the fully-locked solution calculated using the
analysis of Sec.~\ref{Sec_FullLocking} is also shown on Fig.\
\ref{th_pd} as the dotted line. In addition we have performed a
careful numerical scan of the $\alpha-\beta$ plane for $N=1000$ or
$10000$ oscillators with a uniform distribution of full width
unity. These results confirm the analytic predictions, and again
show additional transitions that are inaccessible to our analytic
calculations. The numerics shows that the linear instability from
the unsynchronized state is subcritical over the dashed portion of
the linear instability curves as shown in the figure. Other
saddle-node bifurcations are shown as dash-dotted lines. The
complete phase diagram is quite complicated. Some representative
numerical sweeps are shown in Fig.~\ref{fig_cuts_tophat}.

\subsection{Triangular Distribution}

\label{sec_tri}
\begin{figure}
[tbh]
\begin{center}
\includegraphics[width=3.3in]{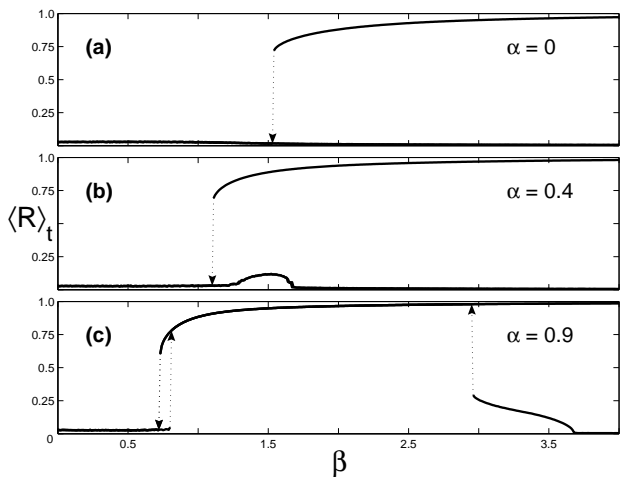}%
\caption{Solutions observed in simulations of $N=1000$ oscillators
having a triangular frequency distribution with $g(0)=1$. The
time-averaged order parameter magnitude $\langle R\rangle_{t}$ is
plotted over a range of $\beta$ at fixed $\alpha$: (a)
$\alpha=0.0$, (b) $\alpha=0.4$,
and (c) $\alpha=0.9$.}%
\label{Fig_tri_cuts}%
\end{center}
\end{figure}

We finally present results for when the oscillators have a triangular
frequency distribution%
\begin{equation}
g(\omega)=\left\{
\begin{tabular}
[c]{cc}%
$(4/w^{2})(w/2-\left\vert \omega\right\vert )$ & for $\left\vert
\omega\right\vert \leq w/2$\\
$0$ & for $\left\vert \omega\right\vert >w/2$%
\end{tabular}
\ \ \ \ \ \ .\ \right.  \label{triangular}%
\end{equation}
This is the case studied in reference \cite{CZLR04}. The
triangular distribution is bounded, and so can have a fully locked
state as for the top-hat distribution, but does not have the
discontinuity in $g(\omega)$ at the edge of the distribution
leading to singular behavior as $\alpha ,\beta\rightarrow0$ for
that distribution. The results are compiled from the linear
stability analysis of the unsynchronized and fully locked state,
as well as numerical investigation, usually of $1000$ oscillators,
as shown in Fig.\ \ref{Fig_tri_cuts}. The integrals
Eqs.~(\ref{integrals})\ needed for the stability analysis of the
unsynchronized state can again be done in closed
form, but the results are too cumbersome to list here.%

\begin{figure}
[tbh]
\begin{center}
\includegraphics[width=3.3in]{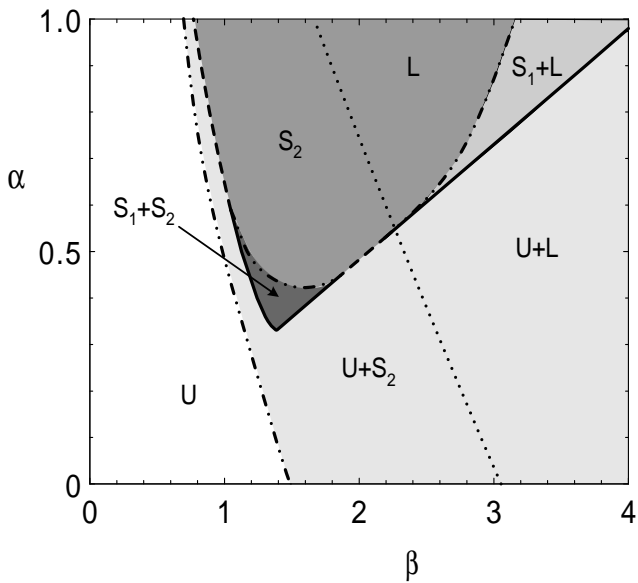}%
\caption{Phase diagram for a triangular distribution of frequencies of full
width $w=2$ such that $g(0)=1$. Symbols and lines are as in Fig.~\ref{th_pd}.}%
\label{tri_pd}%
\end{center}
\end{figure}
The results for $\beta_{c}(\alpha)$ for $g(0)=1$ (a width $w=2$) given by
numerically solving Eqs.~(\ref{perp_integrals}) and (\ref{par_integrals}) are
shown as the full and dashed lines in Fig.\ \ref{tri_pd}. Simulations show that
this transition is subcritical over the dashed
portions. In addition, as in Fig.\ \ref{th_pd},
saddle-node bifurcations identified by jumps in the order parameter magnitude
$R$ in the simulations are shown by dash-dotted curves.

As for the top-hat distribution,
the large $\beta$ limit of the instability region corresponds to an order
parameter frequency \emph{outside} of the band of shifted oscillator
frequencies, so that there is no oscillator with a frequency equal to that of
the order parameter. This seems to be the typical result for a bounded
distribution. For the unbounded Lorentzian distribution on the other hand,
there are some oscillators that lock to the order parameter frequency at the
large $\beta$ instability. However even for this distribution, for most values
of $\alpha$ the order parameter frequency is far in the tails of the
distribution, and there are few oscillators at this frequency. Presumably
the effect of these locked oscillators on the critical $\beta$ is small, so
that the dominant physics at this transition for bounded and unbounded distributions
is not very different. Near the smallest $\alpha$ where the upper and lower stability
boundaries meet, the order parameter frequencies of the two solutions approach
the edge of the band (from either side).

\section{Conclusions}

\label{sec_conclude} In summary, we have analyzed in detail a
model for the synchronization of nonlinear oscillators where the
ordering arises from the reactive coupling between the
oscillators, combined with the nonlinear frequency pulling of the
individual oscillators. Such a model may be a more realistic
description than previous models for a variety of physical systems
where dissipation plays a relatively minor role, for example
high-Q mechanical oscillators and some optical systems. More
generally, the model may give a more complete description of
synchronization where the individual frequencies are internally
tuned in response to a mismatch with other frequencies in the
ensemble, such as might occur in biological systems.

We have presented detailed analytic calculations for the onset of
partial synchronization from the unsynchronized state, as well as
the existence and bounds of the fully locked synchronized state at
large coupling and nonlinearity for the cases of bounded frequency
distributions. The analytical calculations together with numerical
simulations have been used to construct detailed phase diagrams of
the different synchronized states as a function of the two
parameters of the model, the coefficient of the nonlinear
frequency pulling $\alpha$ and the coupling constant $\beta$, for
various frequency distributions. The intersections of the various
synchronized states lead to rich phase diagrams.

There are a number of interesting features of these phase
diagrams. The instability of the unsynchronized state occurs over
a limited range of $\beta$ at fixed $\alpha$, so that the
unsynchronized state regains stability at very large values of the
coupling strength, although a large amplitude synchronized state
also occurs here, often terminated by a saddle-node bifurcation as
$\beta$ is decreased. This large amplitude synchronized state may
also survive down to $\alpha=0$, i.e.\ even in the absence of the
nonlinear frequency pulling. For bounded distributions the state
that develops from the linear instability of the unsynchronized
state as $\beta$ is decreased is synchronized (the order parameter
is nonzero) but there is no frequency locking. The phase diagrams
also show a wide variety of multistability, with one or more
synchronized states and the unsynchronized state coexisting over
various parameter ranges, leading to hysteresis as the parameters
are varied. The multistability may be particularly dramatic for
the large amplitude synchronized states such as displayed in
Fig.~\ref{fig_zig_zag} where many synchronized states coexist,
leading to a band of solutions in the large $N$ limit.

\begin{acknowledgments}
This paper is based upon work supported by the National Science
Foundation under Grant No.~DMR-0314069, HRL Laboratories, LLC
under Contract No.~SR04209, the U.S.-Israel Binational Science
Foundation Grant No.~2004339, and the PHYSBIO program with funds
from the European Union and NATO.
\end{acknowledgments}


\begin{thebibliography}{99}                                                                                               %
\bibitem {BSRW02}M.~Bennett, M.~F.~Schatz, H.~Rockwood, and K.~Wiesenfeld,
Proc.~Roy.~Soc.~Series A \textbf{458}, (2002)

\bibitem {BR02}E.~Buks and M.~L.~Roukes, J.~Microelectromech.~Sys.~\textbf{11}, 802 (2002)

\bibitem {SHSICC03}M.~Sato, B.~E.~Hubbard, A.~J.~Sievers, B.~Ilic, D.~A.~Czaplewski, and H.~G.~Craighead, Phys.~Rev.~Lett.~\textbf{90}, 044102 (2003)

\bibitem {MMS91}P.~C.~Matthews, R.~E.~Mirollo, and S.~H.~Strogatz, Physica D
\textbf{52}, 293 (1991)

\bibitem {W67}A.~T.~Winfree, J.~Theor.~Bio.~\textbf{16}, 15 (1967)

\bibitem {K75}Y.~Kuramoto, in \textquotedblleft International Symposium on
Mathematical Problems in Theoretical Physics\textquotedblright,
ed.~H.~Arakai, Lecture Notes in Physics, \textbf{39}, 420
(Springer, New York, 1975)

\bibitem {ABVRS05}J.~A.~Acebron, L.~L.~Bonilla, C.~J.~P.~Vicente, F.~Ritort,
and R.~Spigler, Rev.~Mod.~Phys.~\textbf{77}, 137-184, (2005)

\bibitem {S88}H.~Sakaguchi, Prog.~Theor.~Phys, \textbf{79}, 39 (1988)

\bibitem {CZLR04}M.~C.~Cross, A.~Zumdieck, R.~Lifshitz, and
J.~L.~Rogers, Phys.~Rev.~Lett.~\textbf{93}, 224101 (2004)

\bibitem {ZZOTSICPC01}M.~Zalalutdinov, A.~Zehnder, A.~Olkhovets, S.~Turner,
L.~Sekaric, B.~Ilic, D.~Czaplewski, J.~M.~Parpia, and H.~G.~Craighead,
Appl.~Phys.~Lett.~\textbf{79}, 695 (2001)

\bibitem{IKARC05} B.~Ilic, S.~Krylov, K.~Aubin, R.~Reichenbach, and
H.~G.~Craighead, Appl.~Phys.~Lett.~\textbf{86}, 193114 (2005)

\bibitem {AEK90}D.~G.~Aronson, G.~B.~Ermentrout, and N.~Kopell, Physica D
\textbf{41}, 403 (1990)

\bibitem{LC03} R.~Lifshitz and M.~C.~Cross, Phys.~Rev.~\textbf{B67}, 134302 (2003)
\end{thebibliography}
\end{document}